\documentclass[fontsize=11pt,a4paper,fleqn]{scrartcl}

\usepackage{a4}
\usepackage{layouts}

\usepackage{amsfonts}
\usepackage{amsmath}
\usepackage{amssymb}
\usepackage{exscale} 

\usepackage{graphicx}
\usepackage[all]{xy}
\usepackage{tikz}
\usetikzlibrary{arrows,matrix,backgrounds,shapes,positioning,fit,chains}
\usepackage{rotating}
\usepackage{caption}
\usepackage{subfig}
\usepackage[noend]{algorithmic}
\usepackage{algorithm}

\usepackage{multirow}
\usepackage{booktabs,dcolumn}
\newcolumntype{d}{D{.}{.}{2.5}}

\usepackage{enumerate}

\usepackage[english]{babel} \usepackage[T1]{fontenc}
\usepackage[ansinew]{inputenc}
\usepackage{url}

\usepackage{comment}
\usepackage{color}
\usepackage[colorinlistoftodos]{todonotes}

\usepackage{makeidx}
\usepackage{natbib}
\setkomafont{sectioning}{\usefont{OT1}{cmss}{c}{n}\selectfont}
\setkomafont{title}{\usefont{OT1}{cmss}{c}{n}\selectfont \huge}
\setkomafont{section}{\LARGE} \setkomafont{subsection}{\Large}
\setkomafont{subsubsection}{\large\itshape}
 \allowdisplaybreaks[4]

\newcommand{\bc}{\begin{center}}
\newcommand{\ec}{\end{center}}
\newcommand{\ben}{\begin{enumerate}}
\newcommand{\een}{\end{enumerate}}
\newcommand{\beq}{\begin{equation*}}
\newcommand{\eeq}{\end{equation*}}
\newcommand{\bea}{\begin{align*}}
\newcommand{\eea}{\end{align*}}
\newcommand{\bi}{\begin{itemize}}
\newcommand{\ei}{\end{itemize}}

\newcommand{\bm}{\boldsymbol}
\newcommand{\bvec}{\left(\begin{array}{c}}
\newcommand{\evec}{\end{array}\right)}
\newcommand{\bmat}[1]{\left(\begin{array}{*{#1}{c}}}
\newcommand{\emat}{\end{array}\right)}

\newcommand{\E}{\mathbb{E}}

\newcommand{\iid} {\operatorname{i.i.d.}}

\newcommand{\package}[1]{\textsf{#1}}

\algsetup{linenodelimiter=}

\title{Penalized Likelihood and Bayesian Function Selection in Regression Models}
\author{Fabian Scheipl, Thomas Kneib \& Ludwig Fahrmeir}

\begin{document}

\maketitle

\begin{abstract}
Challenging research in various fields has driven a wide range of methodological
advances in variable selection for regression models with high-dimensional
predictors. In comparison, selection of nonlinear functions in models with
additive predictors has been considered only more recently.
 Several competing suggestions have been developed at about the same time and
 often do not refer to each other. This article provides a state-of-the-art
 review on function selection, focusing on penalized likelihood and Bayesian
 concepts, relating various approaches to each other in a unified framework. In
 an empirical comparison, also including boosting, we evaluate several methods
 through applications to simulated and real data, thereby providing some
 guidance on their performance in practice.
\end{abstract}

\newpage

\section{Introduction}

Challenging substantive research questions and technological innovations, beginning in molecular
 life sciences, stimulated a revival of methodological and applied research for
selection of influential components in regression models. There exists now a vast
literature on variable selection in linear regression models \begin{equation}\label{eq:lin:mod}
  y_i = \beta_0 + \beta_1 x_{i1} + \ldots + \beta_p x_{ip} + \epsilon_i =
  \eta_i^{\operatorname{lin}} + \epsilon_i, \quad i=1,\ldots,n,
\end{equation}
as well as in generalized linear models or hazard rate models, with
high-dimensional linear predictor $\eta_i^{\operatorname{lin}}$, where $p$ is
(very) large compared to the sample size $n$.

In comparison, corresponding research on function selection in additive models
\begin{equation}\label{eq:add:mod}
  y_i = \beta_0 + f_1(x_{i1}) + \ldots  + f_p(x_{ip}) + \epsilon_i = \eta_i^{\operatorname{add}} + \epsilon_i
\end{equation}
is more recent and still comparably sparse, especially so for
non-Gaussian models with additive predictors
$\eta_i^{\operatorname{add}}$, possibly including further components such as
linear effects and interaction terms.

The following related issues are of
interest in function selection:
 Should a function $f_j(x_j)$ be included in the predictor at all? Because
 functions are usually centered around zero, we therefore want to know whether
 $f_j(x_j)$ differs significantly from zero. Moreover, we are often interested
 whether the effect is linear, that is $f_j(x_j) = \beta_j x_j$ or nonlinear,
 that is, does $f_j(x_j)$ differ significantly from a straight line? Is the
 approximation of the regression function through an additive predictor good
 enough or do we have to incorporate varying coefficient terms, such as $f_1
 (x_1) x_2$, or interactions such as $f_{12}(x_1,x_2)$?

The aims of this article are as follows: We provide a state-of-the-art
survey on recent advances in function selection. Our focus is on penalized least
squares or likelihood methods and on Bayesian concepts, in particular based on
selection indicators and spike-and-slab priors. We sketch - mostly asymptotic - results on model selection
consistency, convergence rates or oracle properties, as far as they are
available. To explore and to illustrate performance in finite sample situations,
we compare and evaluate several methods through application to simulated and
real benchmark data. Naturally, usable software implementation has been a key
criterion for including a selection method into this empirical evaluation.
This is the first empirical comparison of several recent methods for function
selection and provides guidance on their performance in practice.

\sloppy Most frequentist and Bayesian approaches for function selection are based
on or motivated from variable selection in regression models with
high-dimensional linear predictors as in \eqref{eq:lin:mod}. Very popular
penalization methods for variable selection are the Lasso
\citep{Tibshirani:1996}, the SCAD penalty \citep{Fan:Li:2001}, and modifications
such as the adaptive Lasso \citep{Zou:2006}, the groupLasso \citep{Yuan:Lin:2006}
and the groupSCAD \citep{Wang:Chen:Li:2007}. Another branch is boosting
\citep{Buehlmann:Yu:2003, Buehlmann:Hothorn:2007}, which we include as a
further competitor in our empirical analyses since it is applicable to a wide
range of regression models and is implemented in the R package \package{mboost}
\citep{mboost}.

\fussy Bayesian variable selection in linear predictors is often based on
spike-and-slab priors for single regression coefficients and has been most
extensively discussed for the classical linear model \eqref{eq:lin:mod}, see
\citet{George:McCulloch:1993, George:McCulloch:1997}. The basic idea is that each
coefficient $\beta_l$ is modeled as having come either from a distribution with
most (or all) of its mass concentrated around zero (the 'spike'), or from a
comparably diffuse distribution (the 'slab'). A closely related concept is to
introduce selection indicator variables for coefficients being zero or non-zero
\citep{Smith:Kohn:1996}. Instead of placing spike-and-slab priors directly on
regression coefficients, \citet{Ishwaran:2005} assume a spike-and-slab prior for
the variance of (conditionally) Gaussian priors. This concept is conveniently
extended to generalized linear models, see \citet{Fahrmeir:Kneib:Konrath:2010}.
In a recent review paper, \citet{OHara:Sillanpaa:2009} compare several Bayesian
variable selection methods for linear models; see also \citet[Section
4.5.3]{Fahrmeir:Kneib:2011}.

The common feature of penalization methods for function selection is to place a
penalty on certain norms of functional components or associated blocks of basis
function coefficients. The penalty term usually enforces sparseness or smoothness
of regression functions, but both properties are also related to each other
through the choice of the function space. \citet{Lin:Zhang:2006} proposed the
component selection and smoothing operator (COSSO) in additive smoothing spline
ANOVA models, with extensions to exponential family responses in
\citet{Zhang:Lin:2006} and to hazard regression in \citet{Leng:Zhang:2006}.
\citet{Ravikumar:2009} enforce sparsity in additive models by imposing a penalty
on the empirical $L_2$-norm of functional components.
\citet{Meier:Geer:Buehlmann:2009} incorporate an additional smoothness penalty
and \citet{Radchenko:James:2010} impose a further $L_2$-norm penalty to obtain
sparse hierarchical structures in models with two-way functional interactions.
All these methods include only one global penalty parameter to enforce sparseness
of main effects, or a second one to control the amount of smoothness of main
effects or hierarchical structures in two-way interaction models. This leads to a
tendency to oversmooth nonzero functional components in order to set unimportant
functions to zero. Motivated by the adaptive groupLasso, \citet{Storlie:2011}
propose the adaptive COSSO (ACOSSO) to penalize each functional component in
additive models differently so that more flexibility is obtained for functional
components with more curvature while shrinking unimportant components more
heavily to zero. \citet{Zhang:Cheng:Liu:2011} extend the
adaptive COSSO by decomposing nonlinear functions into a linear part and a nonlinear deviation. 
A similar
adaptive groupLasso approach is studied in \citet{Huang:2010}. Instead of penalizing directly the norms of basis function coefficient vectors, leading to groupLasso-type penalties, \citet[for additive
models]{Xue:2009} and \citet[for varying coefficient models]{Wang:Chen:Li:2007},
insert norms into the SCAD penalty, leading to groupSCAD-type penalties.

\sloppy \citet{Belitz:Lang:2008} propose a simple and computationally efficient
method for simultaneous estimation and selection in (generalized) additive models
based on the quadratic P-spline penalty in combination with an extended
backfitting algorithm. In a double penalty approach, \citet{Marra:Wood:2011}
decompose the quadratic smoothing spline penalty for generalized additive models
into two terms corresponding to the null space, e.g. linear functions, and the
penalty range space, e.g. cubic spline deviations from linear functions,
shrinking both of them to zero. A related idea has been suggested in
\citet{Avalos:Grandvalet:Ambroise:2007} for additive models, shrinking linear
functions to zero in Lasso-type fashion.

\fussy While most penalization methods have been primarily developed for additive
models, boosting approaches are also available for more general semiparametric
predictors and non-Gaussian models, see e.g. \citet{Tutz:Binder:2006} and
\citet{Kneib:Hothorn:Tutz:2009}.

Bayesian function selection is mostly based on introducing spike-and-slab priors
with a point mass at zero for blocks of basis function coefficients or,
equivalently, indicator variables for functions being zero or nonzero.
\citet{Wood:Kohn:2002} and \citet{Yau:Kohn:Wood:2003} describe implementations
using a data-based prior that requires two MCMC runs, a pilot run to obtain a
data-based prior for the ``slab'' part and a second one to estimate parameters
and select model components. \citet{Panagiotelis:Smith:2008} combine this
indicator variable approach with partially improper Gaussian priors, as for basis
coefficients of Bayesian P-splines, in high-dimensional additive models. They
suggest several sampling schemes that dominate the scheme in
\citet{Yau:Kohn:Wood:2003}. A more general approach based on double exponential
regression models that also allows for flexible modeling of the dispersion is
described by \citet{Cottet:Kohn:Nott:2008}. They use a reduced rank
representation of cubic smoothing splines with a very small number of basis
functions to model the smooth terms in order to reduce the complexity of the
fitted models, and, presumably, to avoid mixing problems.
\citet{Reich:Storlie:Bondell:2009} consider a Gaussian functional ANOVA model
with (conditionally) conjugate Gaussian process priors for functions. To perform
function selection, they impose a spike-and-slab prior on variances of Gaussian
processes, with a point mass at zero for the spike and a half-Cauchy prior for
the slab. It seems difficult, however, to extend it to non-Gaussian regression
models.

A novel spike-and-slab prior structure for function selection in the broad class
of structured additive regression models for Gaussian and discrete responses is
proposed in \citet{Scheipl:2011}, \citet{Scheipl:Fahrmeir:Kneib:2011} and
implemented in the R package \package{spikeSlabGAM} \citep{spikeSlabGAM}.

\section{Penalized least squares and likelihood-based function selection}\label{sec:penleastsquares}

\subsection{General concept}\label{sec:penleastsquares:general}

Let $\bm y = (y_1,\ldots,y_n)^{'}$ denote the vector of all responses, $\bm \eta
= (\eta_1,\ldots,\eta_n)^{'}$ the vector of corresponding predictor values, and
$\bm f_j=(f_j(x_{1j}),\ldots,f_j(x_{nj}))^{'}$ the vector of function evaluations of a
function $f_j(x_j)$. All penalization methods measure the fit of a
regression model by some loss function $\operatorname{loss}(\bm y,\bm \eta)$ or
equivalently through a utility function $\operatorname{utility}(\bm y,\bm \eta)=
-\operatorname{loss}(\bm y, \bm \eta)$.

The $L_2$-loss
\[
\operatorname{loss}(\bm y, \bm \eta) = ||\bm y - \bm \eta||^2
\] 
i.e., the squared Euclidian distance between $\bm y$ and $\bm \eta$, is
commonly used for additive models $\bm y = \bm
\eta^{\operatorname{add}} + \bm \epsilon$ and extensions including a linear
predictor $\bm \eta^{\operatorname{lin}}$ or interaction terms. Sometimes the
empirical $L_2$-loss $||\bm y - \bm \eta||^2/n$ is considered as a slight
modification. For Gaussian i.i.d. errors the $L_2$-loss coincides with the
negative log-likelihood, up to an additive constant. The $L_2$-loss is also
considered for non-Gaussian errors with distributions with similar shapes, but
then it is only a (negative) quasi-log-likelihood. To improve efficiency, one
may consider a weighted $L_2$-loss, with weights proportional to the inverse
standard deviations.
For non-Gaussian responses with exponential family distributions, in particular binary, categorical or count
responses, the negative log-likelihood \[
\operatorname{loss}(\bm y, \bm \eta) = -\operatorname{l}(\bm y, \bm \eta)
\] is a more natural choice. For Cox-type
hazard regression models or quasi-likelihood models the loss is defined by the
(negative) partial log-likelihood or some pseudo-log-likelihood.

To regularize estimation of high-dimensional regression models and to enforce
sparseness through component selection and/or smoothness of functions, a penalty
function $\operatorname{pen}(\bm \eta)$ is introduced, including one or more
tuning parameters $\lambda$ to control the amount of penalization. For models
with an additive predictor \eqref{eq:add:mod}, the penalty is a sum of penalties
for each function \[ \operatorname{pen}(\bm \eta) = \sum_{j=1}^p
\operatorname{pen}_j (\bm f_j), \] where each penalty term $\operatorname{pen}_j
(\bm f_j)$ includes one (or two) tuning parameter common to all of them, or
separate, in case of tensor product terms possibly multi-dimensional, tuning
parameters $\bm\lambda_j$ for each of them. We distinguish between sparseness
and smoothness penalties. Sparseness penalties are constructed to shrink 
(suitable norms of) functions to zero, deleting ``unimportant'' functions.
Smoothness is - hopefully - obtained as a by-product, e.g. through the choice of
a suitable space of smooth functions. Smoothness penalties explicitly encourage
the fitting of smooth functions. Modifications of well-known smoothing spline
penalized or regression spline approaches are required, however, to encourage
sparseness of predictors. Section \ref{sec:sparsepen} describes sparseness
penalties in more detail, while Section \ref{sec:smoothpen} deals with smoothness
penalties or a combination of both types.

The ultimate goal is to (simultaneously) estimate and select functions $\bm f =
(f_1,\ldots,f_j,\ldots)$ by minimizing the penalized loss \[
\operatorname{loss}(\bm y, \bm \eta) + \operatorname{pen}(\bm \eta) \rightarrow \underset{\bm f \in \mathcal{F}}{\operatorname{min}}
\] where $\mathcal{F}$ is a prespecified suitable function space. Tuning
parameters are assumed to be fixed or known in this minimization problem and are
estimated by minimizing additional criteria, in particular cross-validation,
information criteria such as (modifications of) AIC, BIC, 
and Mallows $\text{C}_p$, or (restricted) marginal likelihood (REML).

In almost all approaches considered in Sections \ref{sec:sparsepen} and
\ref{sec:smoothpen}, the function space $\mathcal{F}_j$ for each function $f_j$
is a linear space defined through basis functions $B_{jk} (x),\; k=1,2,\ldots$,
such as the space of polynomial splines equipped with a B-spline basis. Then
\begin{equation}\label{label3}
  f_j (x) = \sum_k \beta_{jk} B_{jk}(x)
\end{equation}
and the vector $\bm f_j$ of function evaluations can be expressed as
\begin{equation}\label{label4}
  \bm f_j = \bm X_j \bm \beta_j
\end{equation}
with basis function values $B_{jk}(x_{ij})$ as elements of the design matrix $\bm
X_j$. Then the penalized loss can be expressed as \[
\operatorname{loss}(\bm y, \bm \beta) + \operatorname{pen}(\bm \beta)
\rightarrow \underset{\bm  \beta}{\operatorname{min}},
\] where $\bm \beta$ is the vector of all basis
function coefficients. For an additive model \eqref{eq:add:mod} and an
$L_2$-loss, we obtain the penalized least squares criterion

\begin{equation}\label{eq:pls}
  \operatorname{PLS}(\bm \beta) = ||\bm y-\mathbf{1} \beta_0-\sum_{j=1}^p \bm X_j \bm \beta_j||^2 +
  \sum_{j=1}^p \operatorname{pen}_j (\bm \beta_j) \rightarrow \underset{\beta_0, \bm \beta}{\operatorname{min}}.
\end{equation}
with $\mathbf{1} = (1,\ldots,1)^{'}$ denoting a vector of ones.
To guarantee identifiability, appropriate constraints have to be introduced. In
particular, we have to assume non-concurvity of the functions and ensure that
each function estimate is centered around zero.

Many function selection penalties are based on or related to the groupLasso.
\citet{Yuan:Lin:2006} introduced it for linear models as an extension of the
Lasso for selecting groups of coefficients $\bm \beta_j$ associated with factor
variables or a polynomial function of a continuous covariate. The groupLasso
estimator is defined as the solution to
\begin{equation}\label{eq:groupLasso}
  ||\bm y-\mathbf{1} \beta_0-\sum_{j=1}^p \bm X_j \bm \beta_j||^2 + \lambda \sum_{j=1}^p ||\bm \beta_j||_{\bm K_j}\rightarrow \underset{\beta_0, \bm \beta_1,\ldots,\bm \beta_p}{\operatorname{min}}.
\end{equation}
where the design matrix $\bm X_j$ corresponds to the $j$th factor. After
centering covariates and responses, the unpenalized intercept term may be omitted
for simplicity. The norms in the groupLasso penalty are defined by
\[
||\bm \beta_j||_{\bm K_j}=(\bm \beta_j^{'}\bm K_j\bm \beta_j)^{1/2},
\]
where $\bm K_j$ is a $d_j \times d_j$ positive (semi-)definite matrix $\bm K_j$,
$d_j = \operatorname{dim}(\bm \beta_j)$. For $d_j=1,\quad j=1,\ldots,p$, the
groupLasso reduces to the usual Lasso. For $\bm K_j = \bm I_{dj}$, the norm
reduces to the Euclidian norm $||\bm \beta_j||$. To scale for different
dimensions of the coefficient vectors, $\bm K_j = d_j \bm I_{dj}$ may be chosen,
so that $||\bm \beta_j||_{\bm K_j}$ reduces to $\sqrt{d_j}||\bm \beta_j||$.
\citet{Yuan:Lin:2006} discuss shrinkage properties of the groupLasso and suggest
a coordinate-wise algorithm for estimating the coefficients. Solution paths look
similar as for the Lasso: Depending on the shrinkage parameter $\lambda$,
$\hat{\bm \beta}_j$ may be exactly zero and the corresponding group of variables
is removed from the model.

Conceptually, the groupLasso is easily extended to high-dimensional generalized
linear models: The $L_2$-loss has to be replaced by the negative log-likelihood
of a generalized linear model, e.g. a logit model, see
\citet{Meier:Geer:Buehlmann:2008}. However, parameter estimation becomes more
demanding. An implementation can be found in the R-package \package{grplasso}
\citep{grplasso}.

In analogy, the SCAD penalty of \citet{Fan:Li:2001} can be extended to a
groupSCAD penalty by inserting some norm $||\bm \beta_j||_{\bm K_j}$ into the
(scalar) argument of the SCAD penalty. This seems promising, at least from an
asymptotic point of view, because the SCAD selector possesses the oracle
property, in contrast to the Lasso selector.

\subsection{Sparseness penalties}\label{sec:sparsepen}

\citet{Ravikumar:2009} develop function selection for sparse additive models based on the sparseness penalty
\[
\operatorname{pen}(\bm f_j) = \lambda\big(\frac{1}{n}\bm f_j^{'}\bm f_j\big)^{1/2}=\lambda||\bm f_j||_n
\]
where $||\bm f_j||_n$ is the empirical $L_2$-norm, and the functions are assumed
to be centered around zero. Thus, the norm $||\bm f_j||_n$ penalizes deviations from
zero and shrinks function  to zero. The shrinkage parameter $\lambda$ is the same
for all functions. To minimize the PLS criterion, they propose a ``sparse''
backfitting algorithm (SPAM) which works for any smoother, e.g. also for local
linear or kernel smoothers. Thus, smoothness is considered implicitly. Their
approach and parts of the algorithm can be seen as a ``functional'' version of
the groupLasso. This becomes evident if functions are represented through basis
functions. Then $\bm f_j=\bm X_j \bm \beta_j$ and the empirical $L_2$-norm
becomes
\begin{equation}\label{label7}
  \operatorname{pen}(\bm \beta_j) = \lambda ||\bm \beta_j||_{\bm K_j},\quad \bm K_j = \frac{1}{n}\bm X_j^{'} \bm X_j
\end{equation}
which is a groupLasso penalty.

\citet{Radchenko:James:2010} extend the sparse additive model by incorporating
two-way interaction terms $f_{jl}$. They suggest the penalty function
\[
\operatorname{pen}(\bm \eta) = \lambda_1 \sum_{j=1}^p (||\bm f_j||^2 +
\sum_{l=j+1}^p ||\bm f_{jl}||^2)^{1/2} + \lambda_2 \sum_{j=1}^p \sum_{l=j+1}^p
||\bm f_{jl}||,
\] where $||\cdot||$ is the usual $L_2$-norm. The first term
penalizes all model terms, while the second term additionally discourages
interaction terms to enter the model in order to support the hierarchical
structure of main and interaction effects.

\citet{Huang:2010} propose an adaptive groupLasso for additive models, with
functions $f_j (x_j)$ represented through B-splines. After centering the
B-splines $B_{jk}(x)$ by redefining
\[
B_{jk}(x):= B_{jk}-\bar B_{jk},\quad \bar B_{jk}(x)=\frac{1}{n}\sum_{i=1}^n B_{jk} (x_{ij})
\]
(and centering responses around their mean), they suggest to estimate $\bm \beta_j$ through a PLS criterion with the adaptive groupLasso penalty
\begin{equation}\label{eq:agroupL}
  \operatorname{pen}(\bm \beta_j)=\lambda w_j ||\bm \beta_j||.
\end{equation}
While the shrinkage parameter $\lambda$ is estimated through minimizing the BIC,
the weights $w_j$ are determined in a first step via the groupLasso estimator, in
analogy to the adaptive Lasso of \citet{Zou:2006} for linear models. Thus,
another shrinkage parameter has to be estimated in this step. Because the
adaptive Lasso possesses the oracle property, one may expect desirable properties
for the adaptive groupLasso (see Section \ref{sec:propselectors}).

\citet{Xue:2009} and \citet{Wang:Chen:Li:2007} suggest a grouped version of the
SCAD penalty for function selection in additive and varying coefficient models
respectively. Functions $f_j(x_j)$ are again represented through a B-spline
expansion. \citet{Xue:2009} defines the groupSCAD penalty through
\[
\operatorname{pen}(\bm \beta_j) = p_{\lambda}(||\bm \beta_j||_{\bm K_j}),
\]
where $\bm K_j = \bm X_j^{'} \bm X_j/n$ as in \eqref{label7}, while
\citet{Wang:Chen:Li:2007} insert the Euclidian norm $||\bm \beta_j||$ into the
SCAD penalty of \citet{Fan:Li:2001}. It is given by
\[
p_{\lambda}(w)=\begin{cases}
\lambda |w|  & \text{if }|w|\leq\lambda\\
-(w^2-2a\lambda |w|+\lambda^2) & \text{if }\lambda < |w| < a\lambda\\
\frac{(a+1)\lambda^2}{2} & \text{if }|w|>a\lambda,
\end{cases}
\]
which is a quadratic spline with knots at $\lambda$ and $a\lambda$, with $a=3.7$ as a standard option.
Because the SCAD selector in linear models possesses the oracle property, one may
again expect desirable asymptotic properties for groupSCAD selectors (see Section
\ref{sec:propselectors}).

\subsection{Smoothness penalties}\label{sec:smoothpen}

Smoothness penalties are well-known from representing and fitting smooth functions $f_j(x_j)$ in additive and generalized additive models with predictors
\[
\eta_i = \bm x_0^{'}\bm \beta_0 + \sum_{j=1}^p f_j(x_{ij})
\]
through smoothing splines or penalized regression splines. Expressing splines by
their basis function representations $\bm f_j = \bm X_j\bm \beta_j$ in
\eqref{label4}, estimates $\bm{\hat{f}}_j$ and $\bm{\hat{\beta}}_j$ are obtained
as the minimizers of
\[
\operatorname{loss}(\bm y, \bm \beta) + \sum_{j=1}^p \operatorname{pen}_j (\bm \beta_j)
\]
where
\begin{equation}\label{label8}
  \operatorname{pen}_j (\bm \beta_j) = \lambda_j \bm \beta_j^{'}\bm K_j\bm \beta_j = \lambda_j ||\bm \beta_j||_{\bm K_j}^2.
\end{equation}
The loss function is the $L_2$-norm for AMs and the negative exponential family
log-likelihood for GAMs, and $\bm K_j$ are the penalty matrices of smoothing
splines or P-splines measuring roughness of the fitted functions. Although the
smoothness penalties \eqref{label8} look similar to sparseness penalties, there
are three important differences. First, the penalty matrices $\bm K_j$ are only
positive semidefinite and the null space corresponds to functions which are
not penalized. For the most common case of cubic splines, the null space consists of
linear functions. This implies that functions will not be shrunk towards zero
without some modification. Second, instead of the (semi-)norm $||\bm
\beta_j||_{\bm K_j}$, the quadratic norm is used. Thus, $\bm{\hat{\beta}}_j$ is a
Ridge-type but not a Lasso-type estimator. Third, the penalty terms have {\em
separate} smoothness parameters $\lambda_j$ to allow for different degrees of
smoothness of the various functions $f_j(x_j),\, j=1,\dots,p$. In contrast, all
sparsity penalties have one (or two) {\em common} shrinkage parameters, 
leading to a tendency to oversmooth functions.

To circumvent the difficulty that smooth functions will in general not be
estimated as zero even if they are nuisance functions, \citet{Marra:Wood:2011}
impose an additional penalty on the null space. Let
\[
\bm K_j = \bm U_j\bm \Lambda_j \bm U_j^{'}
\]
be the spectral decomposition of the smoothing matrix, with zero eigenvalues in
$\bm \Lambda_j$ corresponding to the null space of $\bm K_j$. Then an extra
penalty can be defined based on $\bm K_j^{*}=\bm U_j^{*}(\bm U_j^{*})^{'}$, where
$\bm U_j^{*}$ is the matrix of eigenvectors corresponding to the zero eigenvalues
of $\bm \Lambda_j$. This allows to impose a double penalty \[
 \lambda_j \bm
\beta_j^{'} \bm K_j \bm \beta_j + \lambda_j^{*} \bm \beta_j^{'}\bm K_j^*\bm \beta_j
\] on each component $f_j$. By estimating $\lambda_j$ and $\lambda_j^{*}$
separately, a function $f_j$ can now be completely removed from the predictor.
For the case of cubic smoothing splines, the first term penalizes deviations from
a straight line and the second term penalizes straight but non-horizontal lines.
This double penalty approach is implemented for GAMs in the R package
\package{mgcv} \citep{mgcv}.

A related idea has been considered by \citet{Avalos:Grandvalet:Ambroise:2007} for
additive models. They impose a Lasso penalty on the coefficients of linear terms.

\citet{Belitz:Lang:2008} develop a computationally very efficient modification
of backfitting for simultaneous component selection and model estimation.
Their approach applies to the very broad class of structured
additive regression, with continuous and discrete response as in generalized
linear models. The additive predictor can include linear terms, smooth functions
as well as interactions, spatial effects for geoadditive regression, 
and random effect terms for clustered or longitudinal data.
Univariate functions are modelled through P-splines, two-way interactions
through corresponding penalized tensor product-splines, spatial components
through Gaussian Markov random fields, and (uncorrelated) random effects are
assumed to be i.i.d. Gaussian. Each of these different types
of ``functions'' $f_j$ is associated with a quadratic penalty \[
 \operatorname{pen}(\bm \beta_j) = \lambda_j ||\bm
\beta_j||_{\bm K_j}^2, \] where $\bm K_j$ depends on the different types of
functions. For univariate smooth
functions, $\bm K_j$ is the usual regression spline penalty matrix, a Markov random
field precision matrix for spatial components, 
and an inverse correlation matrix for Gaussian random effects. Estimation
and selection are carried out by modified backfitting for each of the
components:
At each step, a number of alternative smoothers is computed for a grid of smoothing
parameters, determined through equivalent degrees of freedom, as well as for the
null function, which corresponds to removing the component. The ``best''
alternative, measured through a GCV or AIC score, is selected until the
iterations stop at convergence.

Comparing with sparseness penalties in the previous subsection, it seems natural
to replace quadratic norms $||\bm \beta_j||_{\bm K_j}^2$ in \eqref{label8} by
norms $||\bm \beta_j||_{\bm K_j}$. \citet{Meier:Geer:Buehlmann:2009} consider
additive models \eqref{eq:add:mod}, where (centered) functions are represented
through B-spline basis functions. They introduce a sparsity-smoothness penalty
\begin{equation}\label{eq:sspen}
  \operatorname{pen}(f_j) = \lambda_1\left(||\bm f_j||_n^2 + \lambda_2 \int
  (f_j^{''}(x))^2 dx\right)^{1/2},
\end{equation}
where the first term encourages sparsity, and the second, weighted by the
(common) smoothing parameter $\lambda_2$, measures smoothness through the usual
cubic smoothing spline penalty. With the B-spline representation $\bm f_j=\bm X_j
\bm \beta_j$, one obtains
\begin{equation}\label{label9}
  \operatorname{pen}(\bm \beta_j) = \lambda_1\left(||\bm \beta_j||_{\bm K_j}^2 +
  \lambda_2 ||\bm \beta_j||_{\bm K_j^*}^2\right)^{1/2},
\end{equation}
where $\bm K_j = \bm X_j^{'}\bm X_j/n$ as in
\eqref{label7} and $\bm K_j^*$ is the cubic smoothing spline penalty. As an alternative, they also suggest
\begin{equation}\label{label10}
  \operatorname{pen}(\bm \beta_j) = \lambda_1||\bm \beta_j||_{\bm K_j} + \lambda_2 ||\bm \beta_j||_{\bm K_j^*},
\end{equation}
a weighted sum of the sparseness and the smoothness norm.

\citet{Lin:Zhang:2006} proposed the component selection and smoothing operator
(COSSO) within the framework of functional ANOVA models, decomposing a regression
function $f(x_1,\ldots,x_p)$ into the sum of main effects $f_j(x_j)$, two-way
interaction effects and higher-order interaction effects similar to classical
ANOVA models. Estimation of the components can be based on the theory of
reproducing Hilbert spaces (RKHS), see \citet{Wahba:1990} and \citet{Gu:2002}
for thorough exposures. In practice, however, only main effects or two-way
interaction models are considered. We focus on additive models \eqref{eq:add:mod}
and assume that each function $f_j(x_j)$, $x_j \in [a,b]$, lies in the second
order Sobolev space
\begin{equation*}
\begin{split}
  S_2 = \bigg\{	f :&  f  \text{ and }
  f^{'} \text{ are continuous, } f^{''} \text{ exists and is}\\
  & \text{square-integrable, i.e. } \int_a^b f^{''}(t)dt<\infty\bigg\}.
\end{split}
\end{equation*}
Extensions to higher-order Sobolev spaces and ANOVA decompositions are considered in \citet{Lin:Zhang:2006}.
The norm of a function $f \in S_2$ is defined as
\begin{equation}\label{eq:norm}
  ||f|| = \left[\left(\int_a^b f^2(x)dx\right)^2 + \left(\int_a
  (f^{'}(x)\right)^2 + \left(\int_a^b (f^{''}(x))^2 dx\right)^2\right]^{1/2}.
\end{equation}
The penalty for each main effect is
\[
\operatorname{pen}(f_j) = \lambda ||f_j||,
\]
and the COSSO estimate in additive models \eqref{eq:add:mod} is the minimizer of the PLS criterion
\begin{equation}\label{eq:COSSO}
  \frac{1}{n}\sum_{i=1}^n \left(y_i - \beta_0 - \sum_{j=1}^p
  f_j(x_{ij})\right)^2 + \lambda \sum_{j=1}^p ||f_j||.
\end{equation}
The COSSO penalty is also a sparsity-smoothness penalty, comparable to but different from the sparsity-smoothness
penalty \eqref{eq:sspen} of \citet{Meier:Geer:Buehlmann:2009}: The first term in
\eqref{eq:norm} is the usual (squared) sparsity $L_2$-penalty, while the second
and third term are the squared penalties used in traditional spline smoothing.
However, there is only one common tuning parameter $\lambda$ to control the
amount of sparsity and smoothness.

RKHS-theory shows that minimizers $f_j,\; j=1,\ldots,p$, of \eqref{eq:COSSO} can be represented as
\[
f_j(x) = \sum_{k=1}^n \beta_{jk}(x) K(x,x_{kj})
\]
where $x_{kj},\; k=1,\ldots,n$, are the observed values of the covariate $x_j$,
and $K(s,t)$ is the ``reproducing kernel'' of $S_2$, equipped with the norm
\eqref{eq:norm}. Its explicit form is
\begin{align}
  &K(s,t)=1+k_1(s)k_1(t) + k_2(s)k_2(t) - k_4(|s-t|),\nonumber\\
  \text{with}\;\;\;&k_1(s) = s-0.5,\;k_2(s) = [k_1^2(s)-1/12]/2, \nonumber\\
  &k_4(s) = [k_1^4(s) - k_1^2(s)/2 + 7/240]/24,\nonumber
\end{align}
see \citet[eq. 10.2.4]{Wahba:1990}. Thus $f_j(x)$ has a basis function representation \eqref{label3}, with basis
functions $B_{jk}(x) = K(x,x_{kj})$, and the function penalty $\lambda||f_j||$ defined in
\eqref{eq:norm} can be determined based on this representation as a function of
the regression coefficients to obtain a corresponding PLS criterion.


The COSSO is extended to generalized additive (and interaction) models in
\citet{Zhang:Lin:2006}, replacing the $L_2$-loss by the negative log-likelihood
resulting from exponential family distributions, and to Cox-type hazard rate
models \[
\lambda(t|x_1,\ldots,x_p)=\lambda_0(t)\exp\left(f_1(x_1)+\ldots+f_p(x_p)\right)
\] in \citet{Leng:Zhang:2006}, replacing the $L_2$-loss by the negative partial log-likelihood.

Because of the relationship to the groupLasso, the COSSO will not possess the
oracle property. The common tuning parameter $\lambda$ also leads to a tendency
to oversmooth nonzero function $f_j$ in favor of sparsity. \citet{Storlie:2011}
suggest the adaptive COSSO (ACOSSO) by introducing weights into the COSSO
penalty, resulting in the ACOSSO penalty \[ \lambda \sum_{j=1}^p w_j||f_j|| \quad
\text{or} \quad \lambda\sum_{j=1}^p w_j||\bm \beta_j||_{\bm K_j}, \] in analogy
to the adaptive groupLasso \eqref{eq:agroupL} of \citet{Huang:2010}. The weights
are determined in a first step as \[ w_j = \left(\int \tilde{f}_j(x)
dx\right)^{-\lambda}, \] where $\tilde{f}_j$ is the usual smoothing spline
estimate of $f_j$, and $\gamma > 3/4$, e.g. $\gamma=1$. As \citet{Storlie:2011} show, ACOSSO has
favorable asymptotic properties, including the oracle property, and dominates
COSSO in simulation studies. \citet{zhang2011} extend ACOSSO by 
decomposing nonlinear functions in additive models into a linear effect and a nonlinear
deviation, orthogonal to each other. This allows to decide if a significant effect is linear or nonlinear.

\subsection{Properties of selectors}\label{sec:propselectors}

The following properties are desirable but not always met, even asymptotically.
They are basically the same as for variable selection in models with linear
predictors.

\subsubsection*{Model selection consistency} All influential functions or
components are selected and all non-influential nuisance components are removed
with probability tending to 1 when the sample size tends to infinity. In finite
samples, this property cannot be met exactly, and goodness of selection is
measured through false negative rates, i.e., rates at which influential components
are not selected, and false positive rates, i.e., rates at which spurious nuisance
components are selected. Alternatively, one may of course use true positive and
negative rates.

In models with sparse predictors, where the number of zero components is very
large and may even grow asymptotically faster than the sample size, selection
consistency is also called sparsistency.

Although selection consistency or sparsistency are desirable, they are not always
met. A weaker form is what we call selection sub-consistency: Asymptotically the
true nonzero functions or components are only a subset of the selected
components.

\subsubsection*{Estimation consistency} All functions or components are estimated
consistently, that is, estimates $\hat{f}_j$ converge (in probability) to the
true nonzero or zero functions $f_j$. If functions are represented through some
(truncated) basis function expansion, corresponding groups of coefficients $\bm
\beta_j$ are estimated consistently.

\subsubsection*{Convergence rates and the oracle property} Convergence rates
measure the speed of convergence of consistent estimators. If convergence for
simultaneous selection and estimation in sparse models is as fast as for
estimation of the true (unknown) submodel that only contains the nonzero
components, than the selection and estimation procedure has the oracle property.

For models with high-dimensional linear predictors, the Lasso selects models consistently under regularity assumptions but does not possess the oracle property. The adaptive Lasso
and the SCAD selection procedure possess the oracle property. Therefore, one may expect similar properties for function selection procedures in models with additive predictors based
on the (adaptive) groupLasso or groupSCAD.

Rigorous proofs of asymptotic  properties are only available for additive models.
All authors assume that functions $f_j$ are centered around zero, either by
assuming $\E(f_j(x_j))=0$ for random covariates $x_j$, or by centering functions
empirically, so that $\sum_{i=1}^n f_j(x_{ij})=0$. For simplification, we may
also assume that responses are centered around their mean, so that the
intercept term $\beta_0$ can be omitted.

Most authors relax the conventional assumption of Gaussian i.i.d. errors
$\epsilon_i$ to a certain extent. For example \citet{Huang:2010} assume i.i.d.
errors with $\E(\epsilon_i)=0$ and $\operatorname{Var}(\epsilon_i)=\sigma^2$,
with tail probabilities satisfying $\operatorname{P}(|\epsilon_i|>x)\leq K(\exp-C
x^2)$ for some constants $C>0$ and $K>0$. \citet{Meier:Geer:Buehlmann:2009} and
\citet{Storlie:2011} allow errors to be uniformly sub-Gaussian, that is there
exist constants $K>0$ and $C>0$ such that \[
\underset{n}{\operatorname{sup}}\underset{i=1,\ldots,n}{\operatorname{max}}
\E(\exp(\epsilon_i^2/K))\leq C. \] An important difference is whether the number
$q$ of nonzero functions $f_j\neq0,\;j=1,\ldots,q$, and the number of zero
functions $f_j\equiv0,\;j=q+1,\ldots,p$, are fixed or may grow with a certain
rate with increasing sample size $n$. \citet{Lin:Zhang:2006},
\citet{Storlie:2011} and \citet{Xue:2009} assume that $q$ as well as $p$ (and
therefore the number $p-q$ of zero functions) is fixed. \citet{Huang:2010} assume
that $q$, the number of nonzero functions $f_j\neq0$, is fixed, but $p$ and
therefore the number of zero functions may even grow faster than the sample size.
\citet{Meier:Geer:Buehlmann:2009} consider the case where the numbers of zero and
of nonzero $f_j$'s may both be larger than $n$.

In many applications it seems realistic, that the number $q$ of nonzero
influential functions is fixed while $p$ may grow with $n$ in sparse models.
Therefore, we take a closer look at the assumptions and the results for the
adaptive groupLasso selector of \citet{Huang:2010}. They consider model selection
and estimation in additive models, where each function $f_j$ belongs to a class
$\mathcal{F}$ of smooth functions $f$ on an interval $[a,b]$, say, whose $k$th
derivative $f^{(k)}$ exists and satisfies a Lipschitz condition of order $\alpha
\in (0,1]$, i.e. \[ |f^{(k)}(s)-f^{(k)}(t)|\leq C|s-t|^\alpha \quad \text{for }
s,t \in [a,b]. \] The order of smoothness is defined as $d=k+\alpha$. For
example, the second order Sobolev space $S_2$ is such a class $\mathcal{F}$ with
$d=1+1=2$.

Beyond the assumptions on the errors $\epsilon_i$ already stated, they assume
that the covariate vectors $\bm x_i = (x_{i1},\ldots,x_{ip})^{'},\;
i=1,\ldots,n$, are realizations of a random covariate vector $\bm x$ with
continuous density, with marginal densities $g_j$ of $x_j$ satisfying
$0<C_1<g_j(x)<C_2<\infty$ \'{o}n $[a,b]$ for every $j=1,\ldots,p$.

To select and estimate functions with the adaptive groupLasso, each function
$f_j$ is approximated through a centered B-spline expansion \[
f_j(x_j)=\sum_{k=1}^m \beta_{jk} B_j(x_j), \] where $m$ also grows with $n$ at an
appropriate rate. Finally, the shrinkage parameter $\lambda_0$ of the groupLasso
estimator, needed to estimate the weights $w_j$ of the adaptive groupLasso in a
first step, as well as the shrinkage parameter $\lambda$ of the adaptive
groupLasso penalty, have to increase with the sample size $n$, with rates defined
in a number of theorems and corollaries on selection consistency and convergence
rates of the adaptive (as well as the non-adaptive) groupLasso estimator. In
particular, their Corollary 2 provides sufficient conditions which are easy to
verify and guarantee that the adaptive groupLasso estimator using the groupLasso
as the initial estimator is selection consistent and achieves optimal rate of
convergence, that is it possesses the oracle property. More precisely, Corollary
2 of \citet{Huang:2010} says:

Let the groupLasso estimator with $\lambda_0 \sim \sqrt{n \log(pm)}$ and $m \sim
n^{1/2 d + 1}$ be the initial estimator in the adaptive groupLasso. Suppose that
the number of nonzero components $q$ is fixed and that the above assumptions on
error and covariates hold. If the shrinkage parameter $\lambda \leq
\operatorname{O}(n^{1/2})$ satisfies \[
\frac{\lambda}{n^{(8d+3)/(8d+4)}}=\operatorname{o}(1) \;\; \text{and} \;\;
\frac{n^{1/4d+2}\log^{1/2}(pm)}{\lambda}=\operatorname{o}(1), \] a further
condition restricting its admissible growth rate, then the adaptive groupLasso
consistently selects the nonzero components, that is \[
\operatorname{P}(||\hat{f}_j||_2>0,\; j=1,\ldots,q,\text{ and }||f_j||_2=0,\;
j=q+1,\ldots,p) \rightarrow 1. \] In addition, \[
 \sum_{j=1}^q ||\hat{f}_j -
f_j||_2^2 = \operatorname{O}_p(n^{-2d/(2d+1)}). \]

In particular we obtain the convergence rate $n^{-4/5}$ for $d=2$, which is the
optimal rate for functions with smoothness of order 2 if we had known that
$f_j=0$ for $j=q+1,\ldots,p$ in advance. Thus, the adaptive groupLasso is
oracle-efficient.

For $d=2$, this result is also in agreement with the asymptotic properties of the
ACOSSO \citep{Storlie:2011}, although the admissible function space is different
(second order Sobolev space of periodic functions), the covariate values come
from a tensor product design, and the penalty is different. For appropriate
growth rates of the smoothing parameters, the ACOSSO selector also has optimal
convergence rates and is selection consistent, implying the oracle property.

\citet{Xue:2009} provides model selection consistency and optimal convergence
rates for his groupSCAD selector under somewhat different but qualitatively
comparable assumptions: The density function $g(x)$ of the random covariable $x$
is continuous and bounded away from zero on its (compact) support, the functions
$f_j$ are $d$-times continuously differentiable, the polynomial spline estimators
have degree $d$, and the number of interior knots goes to infinity with the order
$n^{1/2d + 3}$.

\citet{Huang:2010} also show that the (non-adaptive) groupLasso is
sub-consistent, that is the selected functions contain the subset of nonzero
functions with probability tending to one, but the selected set will be larger
than the set of nonzero functions. A similar result on sub-consistency is valid
for the sparseness-smoothness penalty selector of
\citet{Meier:Geer:Buehlmann:2009}.

Asymptotic results as the ones sketched in this subsection provide valuable
insight into the properties of selection and estimation procedures. There
are, however, essential restrictions which may limit the usefulness of
asymptotics of this type in applications:

First, all asymptotic results assume a fixed sequence of penalty parameters, with
appropriately chosen growth rates. In practice, however, penalty parameters
are chosen by data driven procedures such as cross-validation, AIC, BIC, etc.
Second, although the various penalties can also be combined with other loss
functions, in particular (negative) log-likelihoods of high-dimensional
generalized additive models, extensions of existing asymptotic results to these
situations of practical interest are missing and quite challenging. Third, the
assumptions on covariates rule out time trends $f(t)$, which are useful in
predictors for longitudinal data, or spatial effect $f(s)$ in geoadditive
regression models.

\section{Bayesian Function Selection}

\subsection{Bayesian Sparseness and Smoothness Priors}

For Bayesian function selection, the first immediate idea might be to transfer
the penalty-based approaches discussed so far into corresponding prior
formulations. Indeed, if $\operatorname{pen}(\bm f_j)$ is a smoothness or
sparsity penalty for a function, a corresponding prior can always be constructed
via \[
 p(\bm f_j)\propto\exp\left(-\operatorname{pen}(\bm f_j)\right)
\] although the resulting prior may be (partially) improper, i.e., it can not
necessarily be normalized to integrate to one. By construction, large values of
the penalty now correspond to a rather small prior ``probability'' such that a
priori functions with small penalty values are favored. Since the penalties
considered in function selection enforce sparsity and/or smoothness of
functions, the corresponding Bayesian priors also yield an \emph{a priori}
preference for sparse models or smooth function estimates. Quite often, the prior will contain
additional parameters controlling in particular its variability and therefore
governing the impact of the prior on the posterior. These variance parameters
then provide the Bayesian counterpart to the frequentist penalty parameters.

The reformulation of penalties as priors 
establishes an immediate connection between penalized maximum likelihood 
estimates and Bayesian posterior mode estimates. Since the posterior for a model
comprising functions $f_1,\ldots,f_p$ is given by \[
  p(\bm f_1,\ldots,\bm f_p|\bm y) = \frac{p(\bm y|\bm f_1,\ldots,\bm f_p)p(\bm
  f_1)\cdot\ldots\cdot p(\bm f_p)}{p(\bm y)}
\] (assuming \emph{a priori} independence of the functions $\bm f_1,\ldots,\bm f_p$ so
that the joint prior factorizes into the product of the individual priors)
maximizing the posterior is indeed equivalent to maximizing the penalized
log-likelihood \[
 l(\bm f_1,\ldots,\bm f_p) - \sum_{j=1}^p\log\left(p(\bm f_j)\right).
\]

Many commonly applied smoothness penalties are based on basis function
representations of the nonparametric functions such that $\bm f_j=\bm
X_j\bm\beta_j$ and a quadratic penalty \[
 \operatorname{pen}(\bm
 f_j)=\operatorname{pen}(\bm\beta_j)=\lambda_j\bm\beta_j'\bm K_j\bm\beta_j
\] with positive semidefinite penalty matrix $\bm K_j$. The corresponding
Bayesian smoothness prior is then multivariate Gaussian with prior density
\begin{equation}\label{genprior}
 p(\bm\beta_j|\tau_j^2)\propto\exp\left(-\frac{1}{2\tau_j^2}\bm\beta_j'\bm K_j\bm\beta_j\right).
\end{equation}
The penalty matrix $\bm K_j$ can be interpreted as a prior precision matrix
inducing correlation properties. The prior variance $\tau_j^2$ can be linked to
the smoothing parameter of the penalty via
$\lambda_j=\tfrac{1}{2}\tau_j^{-2}$ and therefore has to be interpreted
inversely to the smoothing parameter, i.e., large variances correspond to only
moderate impact of the prior on the posterior, 
while small variances induce a large impact of the prior. In most cases, further
hyperpriors are placed on the variance parameter to enable a data-driven choice
of the prior impact, see for example \citet{Fahrmeir:Kneib:Konrath:2010} or
\citet{Fahrmeir:Kneib:2011}. The most common, conjugate choice would be an
inverse gamma prior $\tau_j^2\sim IG(a,b)$ with $a=b=0.001$ as a default option.

The above construction principle does not only work with smoothness priors but
also with sparseness priors. For example, the Bayesian LASSO \citep{ParCas08}
has been introduced as a Bayesian counterpart to LASSO regression and relies on
the prior specification
\begin{equation}\label{lassoprior}
 p(\beta_j|\lambda) \propto\exp(-\lambda|\beta_j|)
\end{equation}
for an individual regression coefficient. This Laplace prior is exactly the
Bayesian analogue to the absolute value penalty considered in the frequentist
framework. In fact, the equivalence between posterior modes and penalized
maximum likelihood allows to transfer basically all penalties discussed in the
previous section to the Bayesian framework. However, Bayesian inference will
usually rely on Markov chain Monte Carlo (MCMC) simulation techniques that require either closed form full
conditionals or suitable proposal densities. These are typically difficult to
obtain with non-standard priors comprising for example combinations of squared
penalties and absolute value penalties or special penalties like the SCAD. The
most convenient framework is obtained with conditionally Gaussian priors, i.e.,
hierarchical prior specifications where the prior for the basis coefficients is
Gaussian given suitable hyperprior specifications. In fact, quite different
marginal priors for the basis coefficients may arise given specific
hyperpriors,
\citet{	Griffin:Brown:05} and \citet{Polson:Scott:2012} contain overviews of
some sparseness-inducing possibilities and their properties.
For example, the Bayesian LASSO can be formulated as a scale
mixture of Gaussian prior with exponential hyperprior on the variance of the Gaussian distribution in the following hierarchy:
\[
 \beta_j|\lambda,\tau_j^2\sim N(0,\tau_j^2),\qquad \tau_j^2|\lambda\sim
 \text{Exp}(\tfrac{1}{2}\lambda^2).
\]
After integrating out $\tau_j^2$, the regression coefficient
$\beta_j$ has a marginal Laplace prior (\ref{lassoprior}). By treating
$\tau_j^2$ as additional unknown parameters arising from the scale mixture
representation in conditionally Gaussian priors for the regression coefficients,
efficient Gibbs samplers can be constructed if the observation model is also
Gaussian or has a latent Gaussian representation (as for example in the case of
probit models). In all other cases, iteratively weighted least squares proposals
within a Metropolis Hastings algorithm are a suitable alternative, see
\citet{Fahrmeir:Kneib:Konrath:2010} for details.

From a conceptual point of view, the equivalence of priors and penalties could
now be used to implement Bayesian function selection by applying the sparsity
and smoothness penalties from the previous section. For example, the Bayesian
analogue to the group LASSO penalty (\ref{label7}) would induce the prior
\[
 p(\bm\beta_j)\propto \exp\left(-\lambda ||\bm \beta_j||_{\bm K_j}\right),\quad \bm K_j = \bm X_j^{'} \bm X_j
\]
that would replace the Gaussian standard prior (\ref{genprior}). However, the
automatic selection properties of popular sparseness penalties are typically
lost in the Bayesian framework when inference is based on MCMC. There are at
least two reasons for this: First, the marginal posterior medians or means are
easily obtained from MCMC output while the posterior modes are more difficult to obtain
in a sampling-based approach. Unfortunately, posterior means and medians do not
possess the selection properties of the mode unless the posterior is exactly
symmetric around zero. Second, MCMC inference is sampling based and the
inherent sampling variability will almost surely prevent the algorithm from estimating effects to
be exactly equal to zero even if the posterior mean or median should in fact be
zero. However, the Bayesian approach for example to LASSO regularisation also
has a number of remarkable advantages: First of all, the Bayesian inferential
scheme, unlike the frequentist optimisation approaches, does not only provide
point estimates but the complete posterior distribution. As a consequence,
estimation uncertainty is easily addressed and the selection of ``relevant''
effects can be based on suitable measures of this uncertainty. In addition,
Bayesian smoothness and spareness priors can easily be combined with other
complex predictor structures such as nonlinear or spatial effects due to the
modular nature of MCMC algorithms. Finally, there is some empirical evidence
that Bayesian sparseness priors can yield a
considerable improvement in terms of prediction accuracy when being compared to
standard prior structures, see for example
\citet{KneKonFah09,Konrath:Fahrmeir:Kneib:2012} who compare several prior
structures for high-dimensional parametric predictor components in additive
models and additive hazard regression.

\subsection{Indicator Selection Approaches}

Due to the difficulties with Bayesian sparseness priors discussed in the
previous paragraph, Bayesian selection schemes based on indicator variables
have been developed that explicitly include or exclude terms from a model.
For example, model \eqref{eq:add:mod} could be replaced by

\[
 \eta_i = \beta_0 + \gamma_1f_1(x_{i1}) + \ldots + \gamma_pf_p(x_{ip})
\]
where the indicator $\gamma_j\in\{0,1\}$ can be used to include ($\gamma_j=1$)
or remove ($\gamma_j=0$) terms from the model. To further improve Bayesian
function selection, most approaches  discussed in the following use a
decomposition of terms $f_j$ into a simple, parametric part and more complex
deviations. For example, in the context of nonparametric regression where $f_j$
corresponds to the potentially nonlinear effect of a continuous covariate $x_j$,
it makes sense to decompose the total effect into a linear part $x_j\beta_{0j}$
and the deviation from the linear part $\tilde{f}_j(x_j)$, i.e.,
\begin{equation}\label{decomposition}
 f_j(x_j) = x_j\beta_{0j} + \tilde{f}_j(x_j).
\end{equation}
Instead of putting one joint selection indicator on the overall function $f_j$,
one can then assign separate selection indicators to the linear effect and the
nonlinear deviation. As a consequence, one can determine whether the effect of
covariate $x_j$ shall be included at all, whether it can be adequately described
by a linear effect or whether more complex nonlinear modelling is indeed needed.
While the decomposition is easily incorporated in the Bayesian selection
framework by modifying the design matrices of the model before actually starting
estimation, it is usually more difficult to incorporate in frequentist
approaches relying on sparseness or smoothness penalties.

In a Bayesian formulation, the selection indicators $\gamma_j$ are then treated
as additional unknowns that are assigned a Bernoulli prior
$\gamma_j\sim\operatorname{Be(\pi_j)}$ where $\pi_j=P(\gamma_j=1)$ is the prior
inclusion probability of term $f_j$ and additional prior hierarchies may be
placed on $\pi_j$. Incorporating the indicators $\gamma_j$ in an MCMC estimation
scheme allows to estimate posterior inclusion probabilities $P(\gamma_j=1|\bm
y)$ for example by calculating the relative frequency of samples with
$\gamma_j=1$ (although more efficient estimates can be obtained from
Rao-Blackwellization, i.e. by averaging the full conditional probabilities that
are used for proposing a new state of $\gamma_j$ in the MCMC algorithm).
Applying a threshold to the estimated inclusion probability then provides a
means of function selection.

If interest is not on one final model but on incorporating model uncertainty in
the estimation of the functions, Bayesian model averaging can be implemented by
averaging over all samples obtained from the different configurations of
selection indicators. This in fact yields estimates that are weighted averages
of the individual model estimates with weights proportional to the posterior
model probabilities.

\citet{Cottet:Kohn:Nott:2008} consider function selection in double exponential
families where separate predictors can be placed on expectation and 
variance of the responses. Generalized additive models are contained as a
special case and we will restrict our attention to this situation. Nonlinear
effects of continuous covariates are modelled utilizing the RKHS representation
of smoothing splines that exhibits a direct differentiation into a linear
function part and nonlinear deviations as in (\ref{decomposition}) and yields a
zero mean Gaussian smoothness prior for the nonlinear effects $\tilde{\bm f}_j$
with covariances \[
 \operatorname{Cov}(\tilde{f}_j(x_j), \tilde{f}_j(x_j')) =
 \tau_j^2C(x_j,x_j')\qquad\mbox{and}\qquad
 C(x,x')=\frac{1}{2}x^2\left(x'-\frac{1}{3}x\right).
\] To make inference via MCMC numerically tractable, a low rank approximation
based on the spectral decomposition of the covariance matrix is 
employed, leading to 
\begin{equation}\label{eq:Cottet}
 \bm f_j=\bm X_j\bm\beta_j,\quad   \bm\beta_j \sim N(0,\tau_j^2\bm I).
\end{equation} The very small dimension of $\bm\beta_j$ chosen by
\citet{Cottet:Kohn:Nott:2008} in their analyses may also be related to mixing problems associated with larger
blocks of coefficients (as detailed in the next section) but since no
implementation of the approach has been made available by the authors, this
conjecture can not be checked empirically. An inverse gamma prior is assigned to
the smoothing variances $\tau_j^2$ of the nonlinear effects to include
estimation of the required amount of smoothness. To be able to differentiate
between the necessity of linear and nonlinear effects for specific covariates,
each covariate is assigned two indicator variables: one for the linear part of
the effect and one for a completely nonlinear model. By assuming \emph{a priori}
dependence between the indicators such that the indicator for the nonlinear part
is only nonzero if the indicator for the linear effect is nonzero, the model is
made identifiable.

For the selection indicators, \citet{Cottet:Kohn:Nott:2008} consider i.i.d.
Bernoulli priors $\gamma_j\sim B(\pi_j)$ with a uniform prior $\pi_j\sim U[0,1]$
for the inclusion probability. This corresponds to the special case of i.i.d.
$\operatorname{Beta}(a,b)$ priors with $a=b=1$. While a flat prior seems to make
sense intuitively in representing the absence of prior knowledge on the model
complexity, it in fact induces a binomial distribution with success probability
0.5 for the number of terms included in the model such that it favors models of
intermediate size and has prior expectation of models of size $p/2$. As a
consequence, more refined priors have been developed and will be discussed in
more detail in the following in the context of specific proposals.

Instead of cubic smoothing splines, \citet{Panagiotelis:Smith:2008} consider
penalized splines in the spirit of \citet{EilMar96} for modelling nonparametric
effects in a purely additive model. In order to achieve a separation between
linear and nonlinear part of a function, they impose a constrained prior on the
penalized splines that effectively removes the null space from the penalty. When
using penalized splines with second order difference penalty, this means that a
linear constraint is placed on the basis coefficients that restricts the linear
part of the function to be zero. Then, a separate linear effect can be included
in the model and an explicit differentiation is  possible. The linear constraint
in the prior for the penalized spline basis leads to similar constraints in the
posterior but fortunately such a constraint can easily be incorporated when
sampling from Gaussian full conditionals or proposal densities
(see \citep{RueHel05} for corresponding simulation algorithms). As a
beneficial side effect, the constraint also allows to formulate the complete model in terms of proper priors
which is a prerequisite for Bayesian function selection. Due to the explicit
differentiation between linear and nonlinear effects,
\citet{Panagiotelis:Smith:2008} can put separate selection indicators on them
without additional hierarchical requirements. As a prior for the selection
indicators, they do not consider an i.i.d. Bernoulli prior with flat hyperprior
but the prior \[
 p(\bm\gamma) = \frac{1}{p+1}\binom{p}{q_\gamma}^{-1}
\] where $\bm\gamma$ denotes the complete vector of selection indicators, $p$ is
the number of model terms under selection and $q_\gamma$ is the number of
indicators that are non-zero, i.e., the number of terms to be included. This
prior structure implies a uniform prior \[
 p(q_\gamma)=\frac{1}{p-1}
\] on the number of model terms and therefore assigns equal probability to all
possible model sizes. To circumvent sampling problems that have been present in previous
function selection approaches such as \citet{Yau:Kohn:Wood:2003},
\citet{Panagiotelis:Smith:2008} propose a sampler where both the selection
indicator and the basis coefficients are updated simultaneously. However, their
proposal is currently limited to models with Gaussian responses and no
implementation is publicly available.

A related approach to function selection based on categorical instead of binary
selection indicators is proposed in \citet{Sabanes:Held:Kauermann:2011}  to
directly incorporate prior information on the model space in the model
formulation. For each nonlinear effect $f_j$ in the model,
\citet{Sabanes:Held:Kauermann:2011} define a discrete, finite list
$d=\{0,1,\ldots, K\}$ of potential degrees of freedom comprising $df=0$ (no
effect) and $df=1$ (linear effect) as special cases. This can be seen as a categorical extension of the
binary indicators that only allow for inclusion or exclusion of effects and model
the variability of effects given the indicator in a separate step. Instead,
\citet{Sabanes:Held:Kauermann:2011} combine inclusion or exclusion and
variability of the effect estimates in one joint indicator.

The functions $f_j$ are specified in mixed model representation as in
\eqref{eq:Cottet} and a $g$-type prior 
\[
 N(\bm 0, g \bm J_0^{-1})
\]
is specified for the ``fixed effects'' part of the model after marginalizing
with respect to the ``random effects'' part of the resulting mixed model where
$g$ is a scaling factor that is assigned a further hyperprior and $\bm J_0$ is
the prior precision matrix obtained in the marginal mixed model representation.
Posterior inference on the model space is then accomplished by MCMC sampling
where special sampling types are defined to move along the model space. More
specifically, they consider the sampling steps ``Move'',
corresponding to a change in the degrees of freedom of a specific term from the
current value to one of the neighboring values in the prespecified list, 
and ``Swap'', where a pair of terms is selected and the degrees of freedom for the two terms are interchanged.
The latter step is introduced to allow the model to account for collinearity by
swapping terms associated with highly correlated covariates.

The model space for a specific model instance can then be described by a
$p$-dimensional vector $\bm d$ that contains the degrees of freedom for the
various terms. Assuming for simplicity that $K+1$ possible degrees
of freedom have been prespecified for each term (including the special cases
$df=0$ and $df=1$), the prior on the model space is constructed so that all positive degrees
of freedom are equally probable \emph{a priori} for each included model term and
so that the prior on the number of included model terms $q$ is uniform.
This leads to the prior \[ p(\bm d) = \left[(p+1)\binom{p}{q}K^q\right]^{-1}
\] that has the nice feature that for each term the prior inclusion probability
(with any positive value for the degrees of freedom) is $\tfrac{1}{2}$ while
avoiding a preference for models of intermediate size. \citet{Sabanes:Held:Kauermann:2011}
also describe how to adapt the $g$-prior approach to generalized additive models
based on a Laplace approximation. An implementation for both Gaussian and
non-Gaussian responses with binomial or Poisson distribution is provided in the
R-package \package{hypergsplines} \citep{hypergsplines}. The sampling
algorithm is a two-step procedure that first marginalizes over the spline coefficients to sample only model
configurations $\bm d$, and then samples coefficients for models whose
posterior support exceeds a certain threshold.

\subsection{Spike and Slab Priors}

One common problem with the approaches discussed in the previous paragraph is
that the discrete-continuous mixture of priors may lead to considerable
difficulties when trying to achieve satisfactory sampling performance. Quite
often, the mixing properties of the resulting chains are affected by the chain
getting stuck in basins of attraction corresponding to specific selection
configurations. This problem can sometimes be alleviated by considering
marginalized sampling for the indicators after integrating out the corresponding
regression coefficients or by joint updates but these approaches are mostly
restricted to Gaussian responses where -- for example -- marginalization can be
implemented analytically.

As a consequence, a second branch of Bayesian selection approaches replaces the
point mass in zero corresponding to exact selection or exclusion of
effects by a continuous approximation. In this case, the prior for an effect is a mixture
of a very narrow component that basically reduces the effect size to zero (the
so-called spike) and a usual standard prior that is spread out over the
domain of the parameter space (the slab). For example, when selecting single
regression coefficients $\beta_j$, such a prior may be \[
 \beta_j\sim(1-\pi_j) N(0,\tau_0^2) + \pi_j N(0, \tau_1^2)
\] where $\tau_0^2$ is chosen to be a very small constant such that the first
part of the prior approximates a point mass in zero while an inverse gamma prior
may be assigned to the variance of the second prior component $\tau_1^2$. The
approaches based on binary selection indicators discussed in the previous
section can be considered as limiting cases with $\tau_0^2\rightarrow0$ such
that the spike part concentrates in a Dirac measure in zero. However, this
limiting case often induces the same mixing problems and related necessity for
marginalizing out the regression coefficients for sampling the indicator
variable as discussed above.

In the function selection framework, spike and slab priors are more conveniently
placed on the smoothing variances $\tau_j^2$ where the spike could then be an
inverse gamma distribution concentrated close to zero while the slab could be an
inverse gamma prior with an additional hyperparameter. The underlying reasoning
is that when setting the smoothing variance to zero, this induces maximal
penalty and therefore also induces a reduced model. When decomposing the
nonlinear effect into a linear one and nonlinear deviation, setting the
smoothing variance to zero for example induces the deletion of the nonlinear
deviation from the model. There are basically two advantages of placing the
spike and slab prior on the smoothing variance: On the one hand, it avoids the
need to define multivariate spikes and slabs that would be needed for the basis
coefficient vectors and, on the other hand, the original prior structure can be
kept for the basis coefficients and modifications are only required in higher
levels of the prior hierarchy (and therefore also the resulting full
conditionals).

\citet{Scheipl:Fahrmeir:Kneib:2011} propose a very general spike and slab prior
for function selection in structured additive regression models embedded in the
exponential family regression framework where the predictor may not only
comprise linear and nonlinear effects of continuous covariates as in
\eqref{eq:Cottet} but also interaction surfaces, spatial effects or random
effects. Similarly as in \citet{Cottet:Kohn:Nott:2008}, all these terms are
parameterized in the form \eqref{eq:Cottet}. They place the spike and slab
structure on the smoothing variance $\tau_j^2=\gamma_jv_j^2$ with $v_j^2\sim IG(a_j, b_j)$ and the selection indicator following the two-component mixture \[
 \gamma_j\sim (1-\pi_j)\delta_{\nu_0} + \pi_j\delta_{\nu_1}
\] where $\delta_\nu$ denotes a point mass in $\nu$ and $0\le\nu_0\ll\nu_1$ are
constants chosen to scale the variance $v_j^2$ either to a very small value
($\nu_0$ close to zero) or a larger positive value ($\nu_1$ considerably larger
than $\nu_0$). To avoid mixing problems arising with the direct application of
the spike and slab prior structure to coefficient blocks of more than moderate
size, \citet{Scheipl:Fahrmeir:Kneib:2011} propose a multiplicative parameter
expansion where the basis coefficients $\bm\beta_j$ are replaced by
$\bm\beta_j=\omega_j\bm\xi_j$ with a scalar parameter $\omega_j$ that represents
the overall importance of the effect and a vector $\bm\xi_j$ representing the
distribution of the importance across the basis coefficients. The spike and slab
prior structure is then placed on the scalar importance, i.e., $\omega_j\sim
N(0,\gamma_jv_j^2)$, while the components of $\bm\xi_j$ are assumed to be i.i.d.
from the mixture $0.5N(1,1)+0.5N(-1,1)$. As a consequence, the dimension of the
parameter associated with the selection prior is always equal to one, which has
beneficial impact on the mixing behaviour of the MCMC algorithm. Similar as in
the previous approaches, \citet{Scheipl:Fahrmeir:Kneib:2011} apply a mixed model
representation to all model terms that allows to separate for example linear
effects and nonlinear deviations for continuous covariates and therefore allows
to separately include or exclude linear effects and nonlinear deviations. Since
the spike and slab prior is placed on the variances, usual updates for the regression
coefficients can still be used (with some minor modifications due to the
parameter expansion) such that the prior structure can readily be used with all
exponential family response types. The approach is also implemented in the
R-package \package{spikeSlabGAM} \citep{spikeSlabGAM}.

\citet{Reich:Storlie:Bondell:2009} embed their Bayesian function selection
approach in the context of reproducing kernel Hilbert spaces such that both the
basis and the prior precision matrix arise from the reproducing kernel. They
also consider indicators on the smoothing variance, or more precisely on the
smoothing standard deviations $\tau_j$, in the limiting form with a point mass
spike in zero. This leads to a type of zero-inflated prior distribution for the
standard deviations with a point mass in zero and a continuous, positive part.
For the latter, \citet{Reich:Storlie:Bondell:2009} consider a half Cauchy
distribution and propose hyperparameter choices that control the long run false
positive rate of the algorithm. Sampling for the selection indicators is not
performed in a separate step but -- similar as in the proposal by
\citet{Panagiotelis:Smith:2008} via a joint update of the indicator and standard
deviation. Currently, the approach by \citet{Reich:Storlie:Bondell:2009} is
limited to Gaussian responses and also seems hard to extend beyond that setting.

\section{Empirical evaluation}\label{sec:simstudy}

In the following Sections \ref{sec:simstudy:am}, \ref{sec:simstudy:amconc}, and
\ref{sec:simstudy:uci} we compare the performance of a number of approaches, all
of them implemented either in \textsf{R} \citep{R} or \textsf{MATLAB}
\citep{matlab}. To the best of our knowledge, these constitute the
currently available implementations for function selection in additive models.
ACOSSO \citep{Storlie:2011} was left out of this comparison because earlier simulation
studies in \citet{Scheipl:2011} showed it to be not competitive. We did not
include the approach of \citet{Xue:2009}, for which a rudimentary
\textsf{MATLAB} implementation is available, as the available code does not
perform the essential smoothing parameter estimation.
Our empirical evaluation therefore compares:
\begin{itemize}
  \item \texttt{avalos}: Parsimonious additive models described in
  \citet{Avalos:Grandvalet:Ambroise:2007} available as a collection of
  \textsf{MATLAB} scripts. Note that the \textsf{MATLAB} scripts of Avalos et al. were run under the
GNU \textsf{Octave} system \citep{octave}, which is known to be somewhat slower
than \textsf{MATLAB}, so our timings for \texttt{avalos} are pessimistic.
  \item \texttt{hgam}: the high-dimensional additive model approach of
  \citet{Meier:Geer:Buehlmann:2009}, as implemented in R-package \package{hgam}
  \citep{hgam}.
  Smoothing parameters for \package{hgam} are determined via five-fold cross
	validation.
  \item \texttt{hypergsplines}: objective Bayesian model selection with
  penalised splines as described in \citet{Sabanes:Held:Kauermann:2011} and
  implemented in \package{hypergsplines} \citep{hypergsplines}. We used 8 cubic B-spline
basis functions for the designs and ran the sampler for the model indicators
for 10000 iterations, followed by generating 2000 samples from the models
representing 90\% of the posterior mass explored by the model indicator sampling
phase.
  \item \texttt{mboost}: componentwise gradient boosting as implemented in
  \package{mboost} \citep{mboost}. We supply separate base learners for the
  linear and smooth parts of covariate influence.
  We use 10-fold cross validation on the training data to determine the optimal
  stopping iteration (with a maximum of 1000 iterations) and count a baselearner
  as included in the model if it is selected in at least half of the cross-validation runs up to the stopping iteration.
  \item \texttt{mgcv}: the double shrinkage approach for GAM estimation and term selection described in \citet{Marra:Wood:2011},
  as implemented in R-package \package{mgcv} \citep{mgcv}.
  \item \texttt{oraclemgcv}: additive models as implemented in R-package \package{mgcv} based on the true model formula to serve as a
     reference for estimation accuracy.
  \item \texttt{spam}: the approach for sparse additive models by
  \citet{Ravikumar:2009}, available as a collection of \textsf{R}-scripts.
  \item \texttt{spikeSlabGAM}: stochastic search term selection for
  GAMMs as described in \citet{Scheipl:Fahrmeir:Kneib:2011} and implemented in
  \package{spikeSlabGAM} \citep{spikeSlabGAM}. Results for \texttt{spikeslabgam}
  based on the default settings supplied in the package.
 \item \texttt{stepwise}: Stepwise selection as described in
  \citet{Belitz:Lang:2008}, implemented in \textsf{BayesX} \citep{BayesX},
  accessed via \package{R2BayesX} \citep{R2BayesX}.
\end{itemize}
For approaches returning estimated effective degrees of freedom
(\texttt{mgcv, spam}), we count a covariate included as a linear effect if
its estimated effective degrees of freedom are $>10^{-5}$ and as included as a
non-linear effect if its estimated effective degrees of freedom are $>1$.
For approaches returning only the estimated effects (\texttt{avalos, hgam}), we cannot differentiate between linear and non-linear inclusion and count effects
as included both non-linearly and linearly if the norm of the estimated effect
vector is $>0$. For approaches returning estimated posterior inclusion
probabilities (\texttt{hypergsplines, spikeslabgam}), we use probability 0.5 as a threshold for inclusion.

Table \ref{tab:algtable} summarizes the properties and capabilities of
the compared implementations.
\begin{table}[!hb]
\begin{small}
\begin{center}
\begin{tabular}{llccccccc}
 Algorithm & Language & GLM & CIs & Cat. & Lin. & Ran.
 & Non-Sel. & $d>n$\\
  \hline
  \texttt{avalos} &  \textsf{MATLAB}&  & &  &
  $\surd$ &  &  & $\surd$\\
  \texttt{hgam} & \textsf{R} &  $\surd$&  &  &  &
   & & $\surd$\\
  \texttt{hypergsplines} & \textsf{R, C++} &  $\surd$& $\surd$ & $\surd$ &
  $\surd$ &  & & $\surd$\\
  \texttt{mboost} & \textsf{R, C} &  $\surd$&  & $\surd$ & $\surd$ & $\surd$
   & & $\surd$\\
  \texttt{mgcv} & \textsf{R, C} &  $\surd$& $\surd$& $\surd$ &  &
  $\surd$ & $\surd$& \\
  \texttt{spam} & \textsf{R} &  &  &  &
  &  & & $\surd$\\
  \texttt{spikeslabgam} & \textsf{R, C} &  $\surd$& $\surd$& $\surd$ &
  $\surd$ & $\surd$ &$\surd$ & $\surd$\\
     \texttt{stepwise} &  \textsf{R, C++}&
   & $\surd$ & $\surd$ &  &    & & $\surd$\\
\end{tabular}
\end{center}
\end{small}
\caption{Properties and capabilities of
the compared implementations. ``GLM'' indicates whether function selection in
models for non-Gaussian responses (binary, Poisson, etc.) is implemented.
``CIs'' indicates whether any variability estimates for the effects are
returned. ``Cat.'' indicates whether estimation and selection of effects of
categorical covariates is possible. ``Lin.'' indicates whether the implementation
explicitly differentiates between inclusions of linear and non-linear, smooth
effects. ``Ran.'' indicates whether the implementation allows for estimation
and selection of random effects.  ``Non-Sel.'' indicates whether the
implementation allows for a subset of covariate effects that is always included and not under selection.
``$d>n$'' indicates whether the implementation allows for coefficient vectors
whose dimension $d$ is bigger than the number of observations $n$.}
\label{tab:algtable}
\end{table}

\subsection{Simulation study 1: Additive model}\label{sec:simstudy:am}

We simulate data from the following data generating process of a sparse additive model:
\begin{itemize}
  \item We define 4 functions
   \begin{itemize}
    \item $f_1(x) = x$,
    \item $f_2(x) = x + \frac{(2x-2)^2}{5.5}$,
    \item $f_3(x)= -x + \pi\sin(\pi x)$,
    \item $f_4(x)= 0.5x + 15\phi(2(x-.2)) -
        \phi(x+0.4)$, where $\phi()$ is the standard normal density function,
   \end{itemize}
   which enter into the additive predictor.
  \item We define 2 scenarios:
  \begin{itemize}
    \item a ``low sparsity'' scenario: Generate 16 covariates, 12 of which have
        non-zero influence. The true additive predictor is
        \begin{align*}\eta &= f_1(x_1) +
        f_2(x_2) + f_3(x_3) + f_4(x_4) +\\
        & + 1.5(f_1(x_5) + f_2(x_6) + f_3(x_7) +
        f_4(x_8)) + \\
        & +  2(f_1(x_9) + f_2(x_{10}) + f_3(x_{11}) +
        f_4(x_{12})).\end{align*}
    \item a ``high sparsity'' scenario: Generate 20 covariates, only 4 of which have
        non-zero influence and  $\eta = f_1(x_1) + f_2(x_2) + f_3(x_3) +
        f_4(x_4)$.
  \end{itemize}
  \item The covariates are either
   \begin{itemize}
    \item  $\stackrel{\iid}{\sim}U[-2,2]$ or
    \item  from an AR(1) process with correlation $\rho=0.7$.
   \end{itemize}
   \item $y_i= \eta_i + \epsilon_i$, $i=1,\dots,n$ with $\epsilon_i
   \stackrel{\iid}{\sim} N(0, \sigma^2)$
   \item number of observations $n=100, 1000$
   \item signal to noise ratio $\frac{\operatorname{Var}(\eta)}{\sigma^2} = 3,
   10$
   \item We simulate 50 replications for each combination of the various
   settings.
\end{itemize}
This simulation study setup has previously
been used in \citet{Scheipl:2011}. Predictive root mean square error (RMSE) is evaluated on test data sets with
5000 observations. Selection performance, i.e.~how well the different approaches
select covariates with true influence on the response and remove covariates without true influence on the response is measured in terms of accuracy,
defined as the number of correctly classified model terms (true positives and
true negatives) divided by the total number of terms in the model.
For example, the full model in the ``low sparsity'' scenario has 32 potential terms under selection
(linear terms and basis expans ions/smooth terms for each of the 16 covariates),
only 21 of which are truly non-zero (the linear terms for the first 12 covariates plus the 9 basis expansions of the
covariates not associated with the linear function $f_1(x_1)$).
Accuracy in this scenario would then be determined as the sum of the correctly included model terms plus the correctly
excluded model terms, divided by 32.

\begin{figure}[htbp]
\begin{center}
\includegraphics[width=\textwidth]{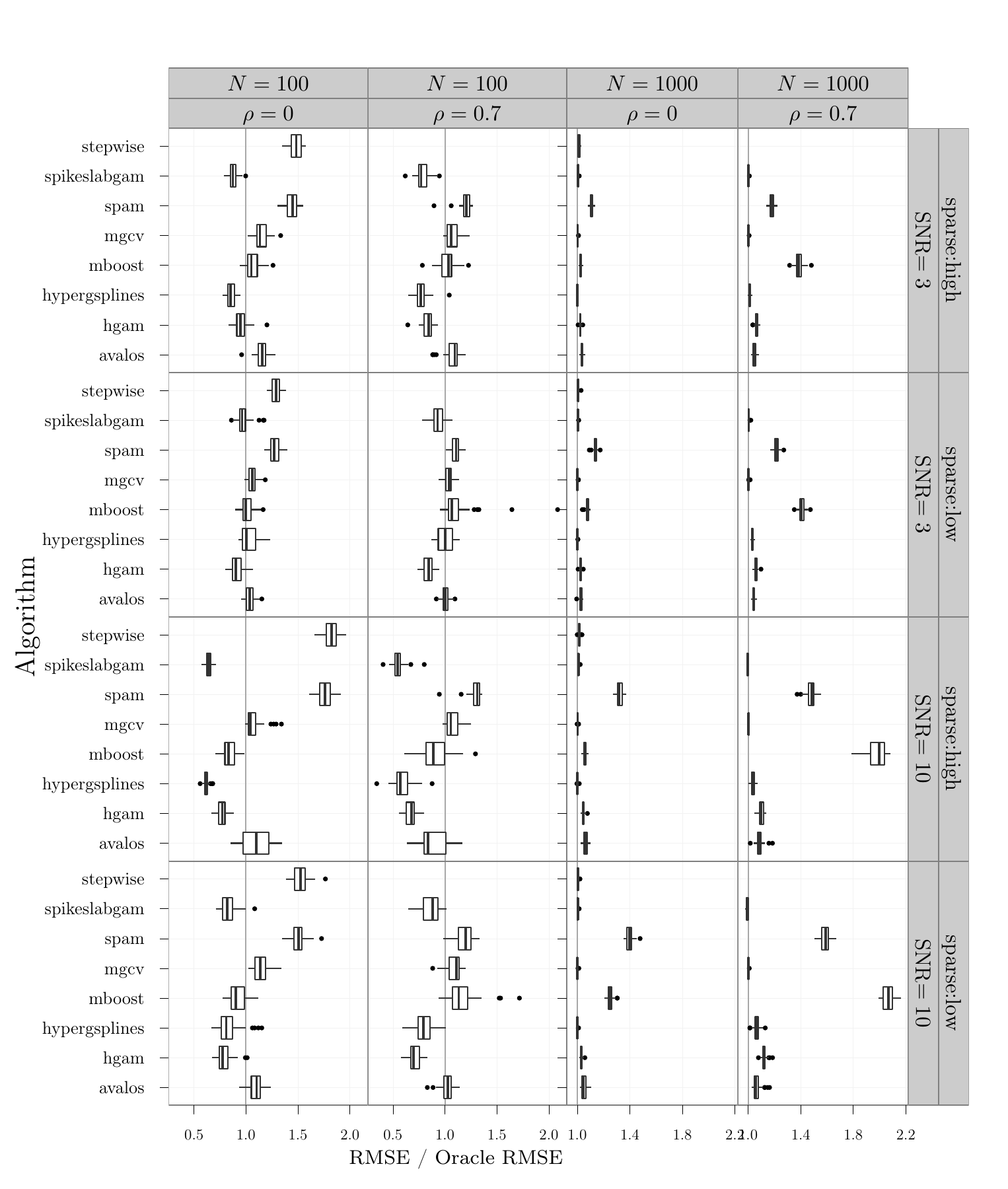}
\caption{Relative RMSE for the different
settings and algorithms. Rows for the different combinations of sparsity and
signal-to-noise ratio, columns for the different combinations of number of
observations $N$ and correlation $\rho$ of the covariates. Vertical gray line
denotes a relative RMSE of 1, i.e. parity with the \package{mgcv}-estimate of
the ``true'' model. Note that results for \texttt{stepwise} are not shown for
correlated responses since it did not return reasonable predictions. Also note that
\texttt{hypergsplines} did not return results for some data sets and only available fits are shown here.}
\label{fig:amrelrmse}
\end{center}
\end{figure}

\begin{figure}[htbp]
\begin{center}
\includegraphics[width=\textwidth]{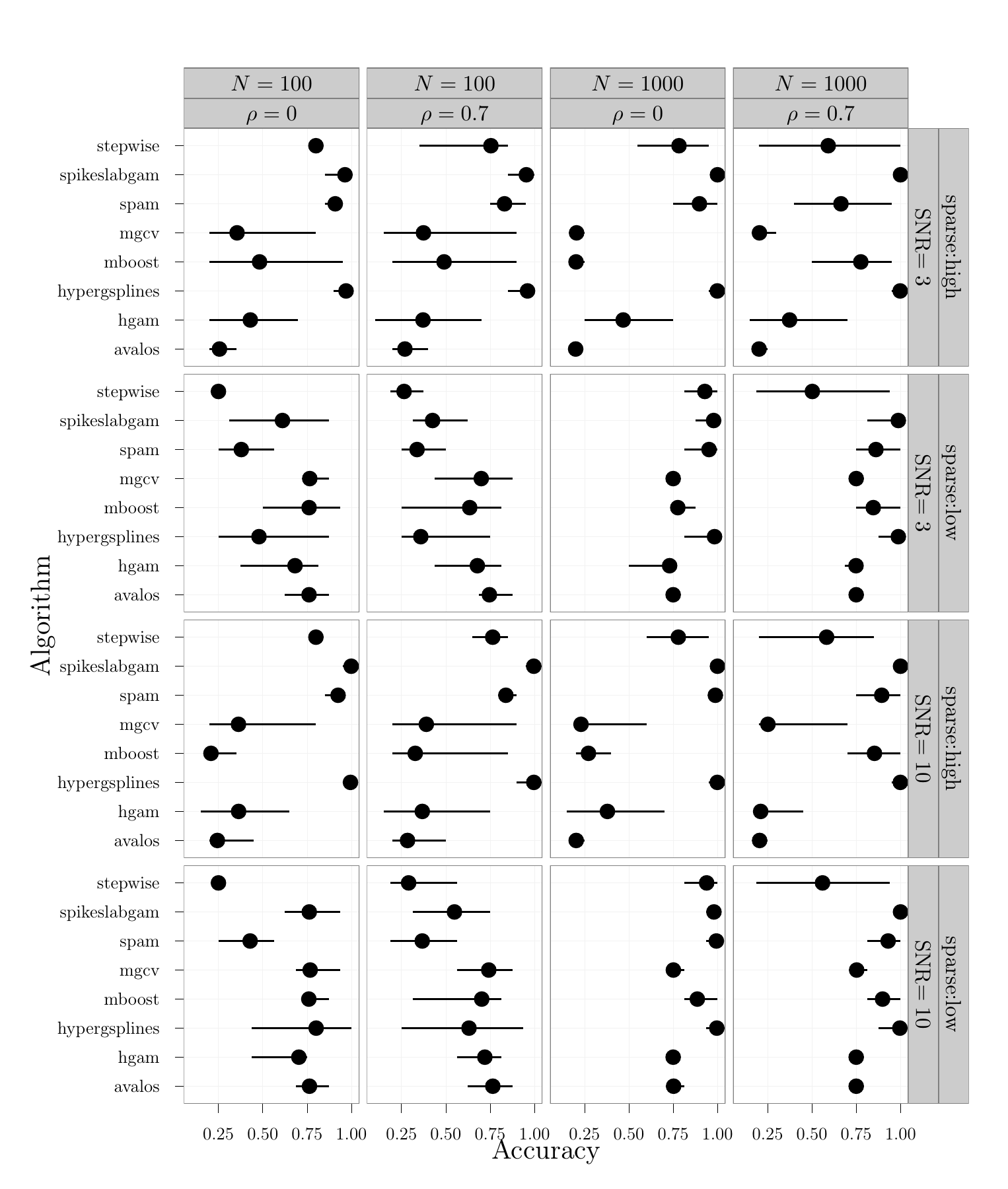}
\caption{Proportion of correctly in- or excluded covariates for the different
settings and algorithms. Rows for the different combinations of sparsity and
signal-to-noise ratio, columns for the different combinations of number of
observations $N$ and correlation $\rho$ of the covariates. Dot denotes mean
accuracy, lines stretch from minimum to maximum
accuracy observed over the 50 replications. Note that results for \texttt{stepwise} are not shown for
correlated responses since it did not return reasonable predictions. Also note that
\texttt{hypergsplines} did not return results for some data sets and only available fits are shown here.}
\label{fig:amacc}
\end{center}
\end{figure}

\begin{figure}[htbp]
\begin{center}
\includegraphics[width=\textwidth]{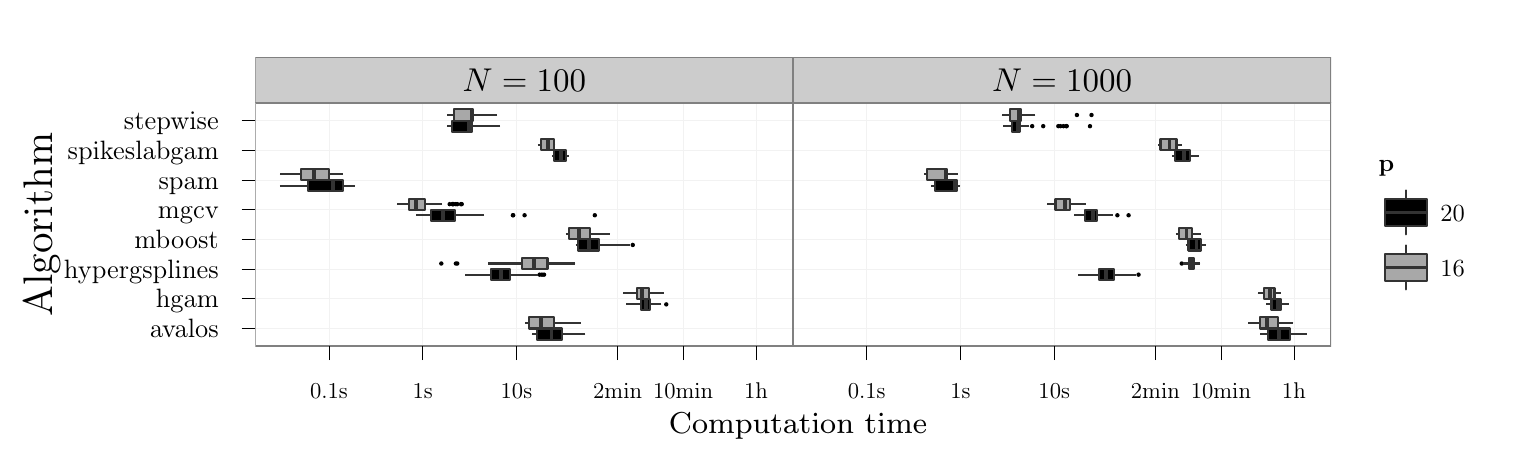}
\caption{Computation times. Boxplots are color-coded for the dimensionality of
the covariate vector. $p=20$ corresponds to the sparse setting with 4 no-zero
and 16 zero effects and $p=16$ corresponds to the non-sparse setting with 12
non-zero and 4 zero effects. Horizontal axis on $\log_{10}$-scale.
Note that results for \texttt{stepwise} do not include timings for
correlated responses since it did not return sensible fits there. Also note that
\texttt{hypergsplines} did not return results for some data sets and only
timings for the available fits are shown here.}
\label{fig:amt}
\end{center}
\end{figure}

Figure \ref{fig:amrelrmse} shows relative RMSE for the different
settings and algorithms. Relative root mean square errors are always relative to
the error of the ``true'' model estimated with \package{mgcv}
(\texttt{oraclemgcv}). Note that results for \texttt{stepwise} for correlated
covariates are omitted as they were orders of magnitude larger than the rest of the results.
Figure \ref{fig:amacc} shows the accuracy of the
variable selection performance in terms of the proportion of correctly excluded
or included effects for the different
settings and algorithms.
Figure \ref{fig:amt} shows computation times for the different algorithms and
settings

Most implementations were very stable, returning useful results for all 7200
model fits. Notable exceptions were \texttt{stepwise}, which failed to return
reasonable fits for any replication of the settings with correlated covariates,
and \texttt{hypergsplines}, which did not return predictions for 37 [47] of the
50 replicates for $N=100$, SNR$=3$ and low sparsity and correlation 0 [0.7] and
also failed in 25 replications for $N=100$, SNR$=3$, low sparsity and
correlation 0.7. Graphs and discussed results are based on  all
available results. Using only replicates for which \texttt{hypergsplines} also
returned predictions would not change the results qualitatively.

Results for relative RMSE (c.f.~Figure \ref{fig:amrelrmse})):
\begin{itemize}
  \item for small $N$, low SNR (4 panels on top left),
  \texttt{spikeslabgam} and \texttt{hgam} consistently beat the error of the
  \texttt{oraclemgcv} benchmark -- \texttt{hypergsplines} does too, but only for
  the sparse setting.
  \item  for small $N$, high SNR (4 panels on bottom left),
  \texttt{spikeslabgam, hgam} and \texttt{hypergsplines} consistently beat the error of the \texttt{oraclemgcv}  benchmark.
  	\texttt{mboost} and \texttt{avalos} also do well for the sparse setting.
  \item  for larger $N$ (two rightmost columns), \texttt{spikeslabgam, mgcv}
  and, to a lesser extent, \texttt{hypergsplines} achieve performance very close
  to that of \texttt{oraclemgcv}. \texttt{stepwise} does too, but only for
  non-correlated responses. \texttt{spam} and \texttt{mboost} do (much) worse
  than \texttt{oraclemgcv} across the board for large $N$.
\end{itemize}
Results for selection accuracy (c.f.~Figure \ref{fig:amacc})):
\begin{itemize}
  \item with the notable exception of \texttt{mboost} in high $N$, high sparsity
  settings and \texttt{stepwise}, selection accuracy is robust against
  correlated covariates for all algorithms.
  \item for $N=1000$, \texttt{spikeslabGAM, hypergsplines}, and to a lesser
  extent \texttt{spam} as well have almost perfect accuracy across the board.
  \item for $N=100$, we see a clear split into algorithms that tend to fit very
  sparse models (\texttt{spikeslabGAM, spam, hypergsplines}) and consequently do
  well in the sparse settings and more liberal algorithms (\texttt{mgcv,
  mboost, hgam}) that do relatively better in the non-sparse settings.
  In the less noisy non-sparse settings, \texttt{spikeslabGAM} and
  \texttt{hypergsplines} also do well.
\end{itemize}
Computation times are displayed in Figure \ref{fig:amt}. The
times of \texttt{avalos} and \texttt{hgam} scale very badly for larger data
set size, while computation times for other methods that exhibit better or at
least competitive selection and estimation performance
(i.e., \texttt{spikeslabGAM, hypergsplines}, and, to a lesser extent,
\texttt{mboost}) remain feasible. As \texttt{hypergsplines} first samples only
the model indicators and then samples from a selection of models with strong
posterior support to arrive at model-averaged effect estimates, it achieves a
significant speed-up in the sparse case ($p=20$) where the models to be sampled
in the second step are very small. All other methods are slower for $p=20$ than for $p=16$.
Note that timings for \texttt{spikeslabGAM, hypergsplines} and \texttt{mboost}
are (very) pessimistic since they do not make use of the parallelization
features available for these 3 algorithms (parallel MCMC chains for
\texttt{spikeslabGAM}, parallel computation via \textsf{OpenMP} for
\texttt{hypergsplines}, parallel bootstrap iterations to determine the stopping iteration for \texttt{mboost}).
\begin{figure}[!ht]
\begin{center}
\includegraphics[width=\textwidth]{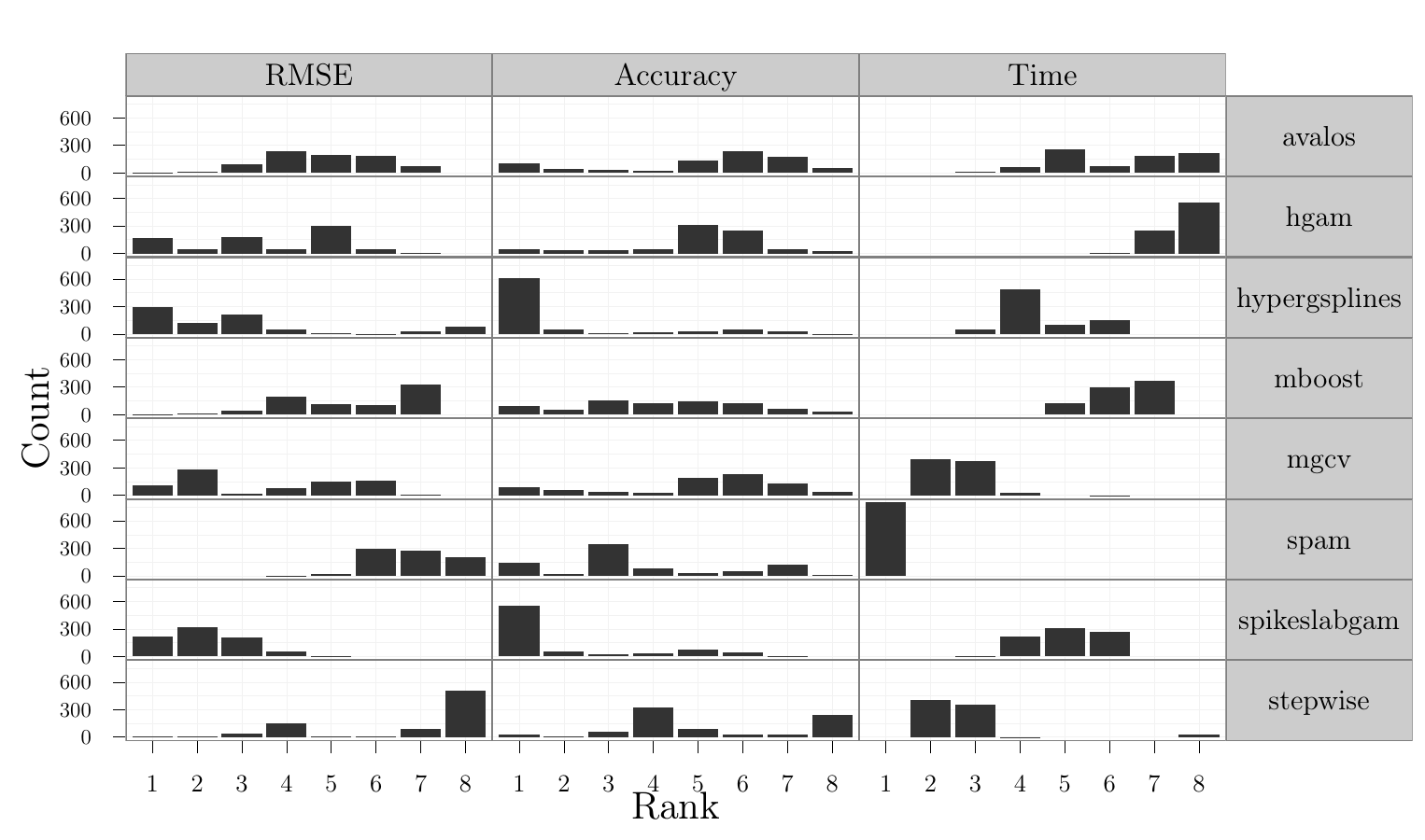}
\caption{Rank distributions achieved by the 8 different algorithms for the AM
simulation. Ties broken by assigning minimum rank to all involved. }
\label{fig:amrank}
\end{center}
\end{figure}
Figure \ref{fig:amrank} shows the distribution of ranks achieved on each
replicate of each setting by the 8 different algorithms in order to directly compare relative
performance in terms of RMSE, selection accuracy and required computation time.
\texttt{spam} and \texttt{stepwise} are fast, but not competitive in
terms of estimation or selection.  \texttt{spikeslabgam} and
\texttt{hypergsplines} deliver the best estimation and selection performance,
with slightly more best performances in shorter time for \texttt{hypergsplines},
but more stable results for \texttt{spikeslabgam}. \texttt{hgam} often delivers
good estimates, but it is very slow. The best compromise in terms of
accurate estimation and computation time seems to be offered by \texttt{mgcv},
but the selection accuracy is below average. Both \texttt{mboost} and
\texttt{avalos} show intermediate results in all three performance dimensions we
consider. Note that the threshold of $>10^{-5}$ effective degrees of freedom for
inclusion of a term that we use for \texttt{spam} and \texttt{mgcv} is very low,
leading to comparably less sparse models and more falsely included covariates
than for most other methods, especially so for \texttt{mgcv}. In that sense,
results for the selection accuracy of  \texttt{mgcv} and, to a lesser extent,
\texttt{spam} are somewhat pessimistic, especially in sparse settings.

\subsection{Simulation study 2: Additive model with
concurvity}\label{sec:simstudy:amconc}

To investigate the various methods' robustness to concurvity, we use a similar
data generating process as the one used in the previous Subsection:
\begin{itemize}
  \item We define functions $f_1(x)$ to $f_4(x)$ as in the data generating process for the previous Subsection.
  \item We use $p=10$ or $p=20$ covariates: The first 4 are associated with
  functions $f_1$ to $f_4$, respectively, while the remainder are ``noise''
  variables without contribution to the additive predictor.
  \item covariates $x$ are $\sim U[-2, 2]$
  \item we distinguish 3 scenarios of concurvity:
  \begin{itemize}
    \item in scenario 1, $x_4 = c\cdot g(x_3) + (1-c)\cdot u$,
     i.e., two covariates with real influence on the predictor are functionally
    related.
    \item in scenario 2, $x_5 = c\cdot g(x_4) + (1-c)\cdot u$,
     i.e., a ``noise'' variable is a noisy version of a function of a covariate
    with direct influence.
    \item in scenario 3, $x_4 = c\cdot g(x_5) + (1-c)\cdot u$,
     i.e., a covariate with direct influence is a noisy version of a ``noise''
    variable.
  \end{itemize}
  where $g(x)= 2\Phi(x, \mu=-1, \sigma^2=0.16) + 2\Phi(x, \mu=1, \sigma^2=0.09)
  - 4\phi(x) -2$ with $\Phi(x,\mu,\sigma^2)$ defined as the cdf of the
  respective Gaussian distribution, $\iid$ standard normal variates $u$, and the
  parameter $c$ controlling the amount of concurvity:
  $c=1$ for perfectly deterministic relationship, and $c=0$ for independence. In our simulation, $c = 0, .2, .4, .6, .8, 1$.\\
  \item we use signal-to-noise ratio SNR$=1, 3$.
  \item we use $n=100$ observations for the training data set.
  \item we simulate 50 replications for each combination of the various settings.
\end{itemize}
Predictive MSE is evaluated on test data sets with 5000 observations.
A similar simulation study setup has previously been used for comparing a
different set of selection and estimation algorithms in
\citet{Scheipl:Fahrmeir:Kneib:2011}. \texttt{stepwise} was omitted from this comparison due to its dismal performance
for correlated covariates (see previous subsection). Note that \texttt{mgcv} did
not return any fits for $p=20$ as it cannot be used to estimate models with more
coefficients than observations.

\begin{figure}[htbp]
\begin{center}
\includegraphics[width=\textwidth]{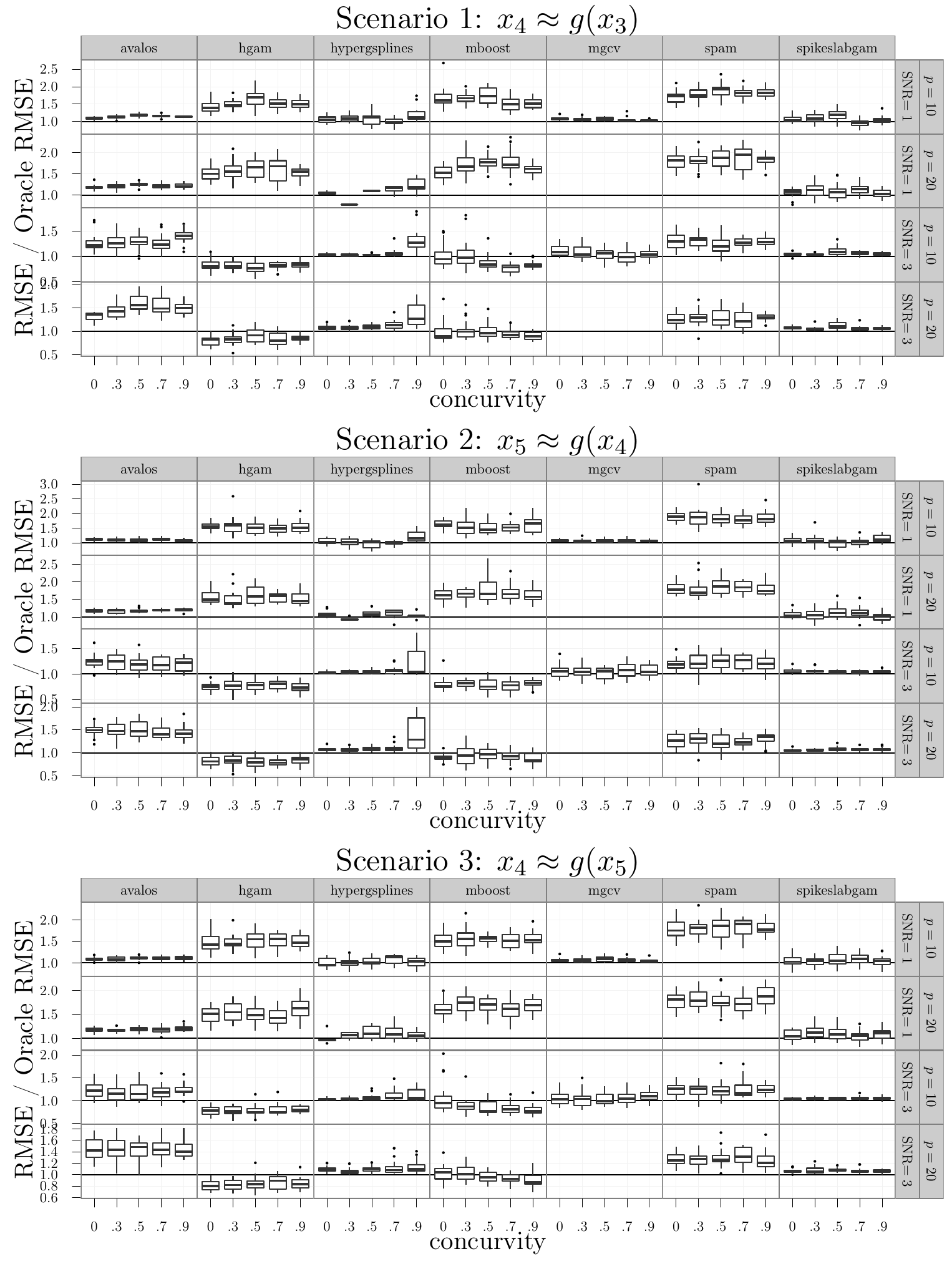}
\caption{Boxplots of prediction RMSE divided by prediction RMSE of the ``true''
model fit by \texttt{mgcv}. Horizontal line denotes relative RMSE of 1, i.e.
parity with the ``oracle'' model. Columns correspond to the different
algorithms, rows correspond to the 4 combinations of number of potential predictors $p$ and
signal to noise ratio. Results for \texttt{mgcv} for $p=20$ are missing as it
cannot deal with models with more coefficients than observations.}
\label{fig:amconcallrmse}
\end{center}
\end{figure}

Figure \ref{fig:amconcallrmse} shows relative predictive RMSE for $y$ for the 7
algorithms and the three scenarios under different concurvity, number of
potential predictors $p$ and signal-to-noise ratio.

Results for relative RMSE under concurvity (c.f.~Figure
\ref{fig:amconcallrmse}):
\begin{itemize}
  \item Overall estimation accuracy of $E(y)$ seems fairly robust
  against increasing concurvity and different number of noise terms for all the
  methods we consider here. Relevant increases in the relative error as
  concurvity increases are visible only for \texttt{hypergsplines} in Scenarios
  1 and 2.
  \item \texttt{hypergsplines, mgcv} and \texttt{spikeslabgam} achieve
  performance almost equivalent to or better than  the ``true''
  model estimates across the board.
  \item \texttt{mboost} and \texttt{hgam} dominate the other methods and the
  ``true'' model estimates for higher signal-to-noise ratios (lower two lines in
  each figure), but are not competitive for more noisy data (upper two lines in
  each figure).
\end{itemize}

\subsection{Benchmark study on binary response data}\label{sec:simstudy:uci}

We investigated the performance of a subset of algorithms on four real data sets
for binary classification taken from the UCI Machine Learning Repository
\citep{Frank:Asuncion:2010}: the ``Musk (Version 1)'' (\texttt{musk}), ``Liver Disorders'' (\texttt{liver}),
``Connectionist Bench (Sonar, Mines vs. Rocks)'' (\texttt{sonar}) and ``Pima
Indian Diabetes'' (\texttt{pima}) datasets. These datasets were picked
since they do not contain any categorical predictors for which none of the
algorithms except \texttt{mboost} and \texttt{spikeslabgam} can perform
selection. Of the implementations described in the previous section,
only  \texttt{hgam}, \texttt{hypergsplines}, \texttt{mboost}, \texttt{mgcv} and
\texttt{spikeslabgam} can deal with non-Gaussian data, so our comparison is
limited to those five approaches.

Table \ref{tab:ucidata} summarizes properties of the four datasets.
\begin{table}[!hb]
\begin{small}
\begin{center}
\begin{tabular}{l rrdd }
Name & $n$ & $p$ & \multicolumn{1}{c}{$n/p$} & \multicolumn{1}{c}{Balance} \\
\hline
\texttt{liver} & 345 & 7 & 49& 1.4\\
\texttt{musk} &  467 & 167 & 2.8& 1.3\\
\texttt{sonar} & 208 & 61 & 3.4& 1.1\\
\texttt{pima} &  768 &  9& 85& 1.9\\
\end{tabular}
\end{center}
\end{small}
\caption{Properties of the four benchmark datasets. $n$ denotes the number of
observations, $p$ denotes the number of potential covariates, ``Balance'' is
the ratio of the number of cases in the larger class divided by the number of
cases in the smaller class, i.e., it is 1 if the data set has equal number of
successes (ones) and failures (zeros).}
\label{tab:ucidata}
\end{table}
Note that only \texttt{musk}, and, to a lesser extent, \texttt{sonar}, are truly
high-dimensional variable selection problems and that only \texttt{pima} is
somewhat unbalanced with a ratio of about 2:1 between cases in the bigger class
and cases in the smaller class.
\begin{figure}[!ht]
\begin{center}
\includegraphics[width=\textwidth]{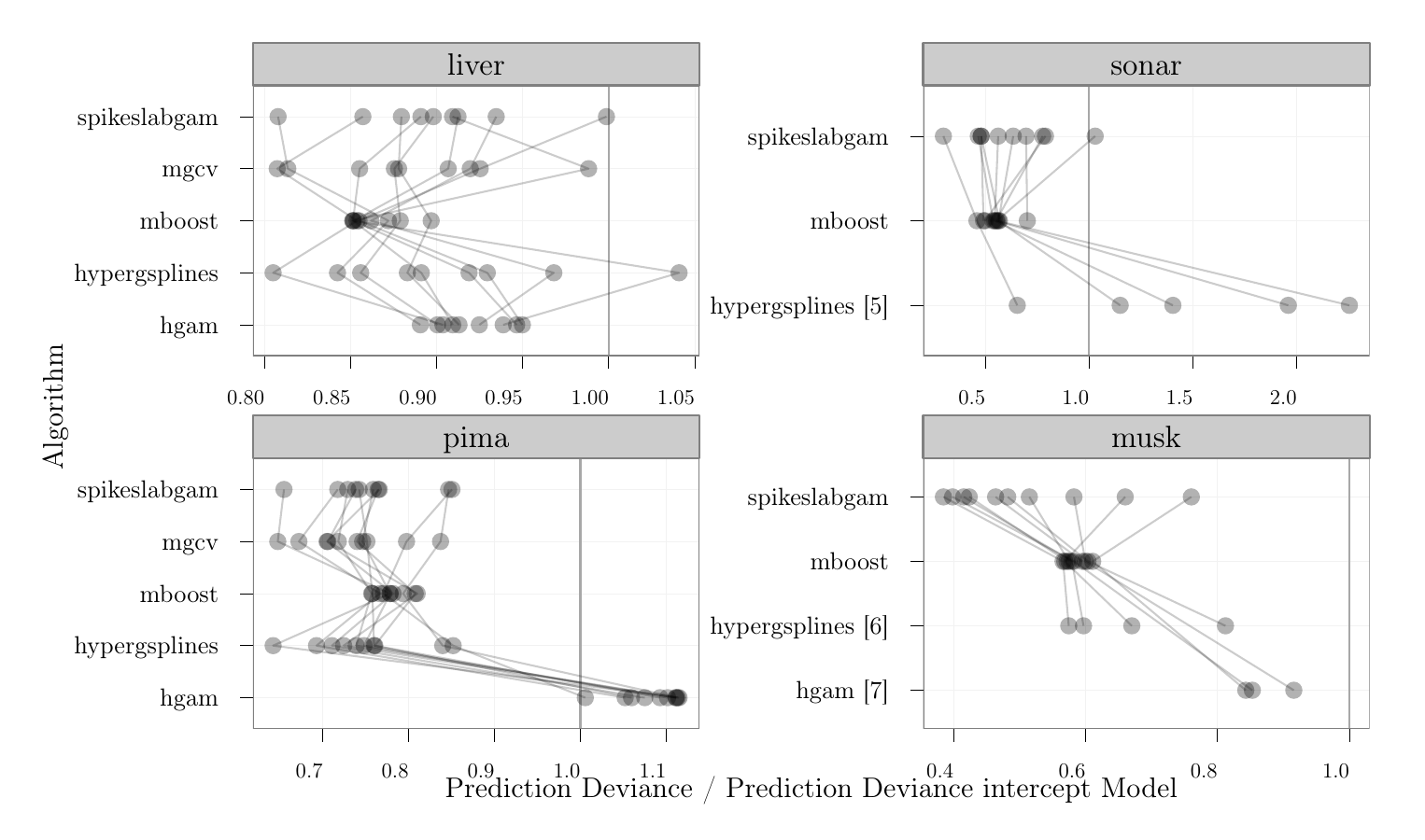}
\caption{Dotplots of relative predictive deviance on the test data sets.
Relative predictive deviance is defined as twice the negative
log-likelihood evaluated on the test data sets divided by the predictive deviance of the
intercept model on the training data, i.e., lower is better and values above 1
indicate predictions worse than those of a simple intercept model on average and indicate substantial
overfitting. Results on the same test set connected by lines. Numbers in
brackets after algorithm names indicate the number of times the algorithm failed
to fit a model and/or return predictions. For the datasets in the right column,
 the full models have more coefficients than observations.}
\label{fig:ucidev}
\end{center}
\end{figure}
\begin{figure}[!ht]
\begin{center}
\includegraphics[width=\textwidth]{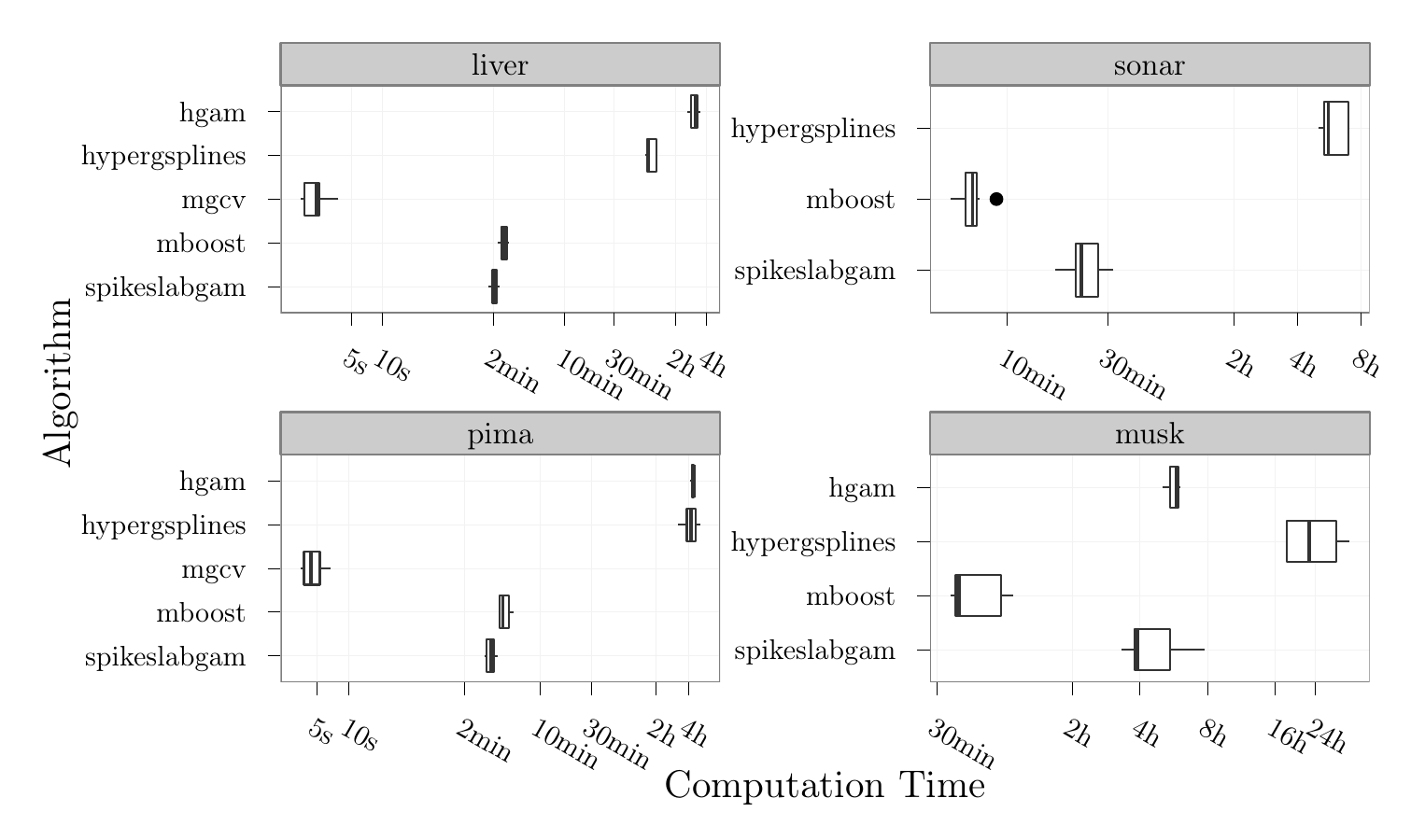}
\caption{Boxplots of computation time for binary classification data sets in
minutes. Time axis on binary log-scale.}
\label{fig:ucit}
\end{center}
\end{figure}

For each of the four datasets, we generated ten bootstrap training data sets and
evaluated predictive performance for the fitted additive logistic models
in terms of minus twice the log likelihood (deviance) on the out-of-bag samples.
To avoid extrapolation issues for the smooth terms, all training data sets
contain those observations with the minimal and maximal values of the potential
covariates for each data set.
Figure \ref{fig:ucidev} depicts the ratio between the predictive deviance
of the fitted models and a simple logistic intercept model fitted on the
training data, i.e., lower is better. Values above 1
indicate predictions worse than those of the simplest conceivable model on
average and indicate substantial overfitting. Note that \texttt{mgcv} can not
fit the \texttt{musk} and \texttt{sonar} datasets since the full model would
have more coefficients than observations. \texttt{hgam} did not return any
predictions for \texttt{sonar} due to numerical problems. For the
remaining data sets, numbers in brackets after algorithm names indicate the
number of times the algorithm failed to fit a model and/or return predictions.
Results for the binary classification benchmark (c.f.~Figures \ref{fig:ucidev},
\ref{fig:ucit}) show that:
\begin{itemize}
  \item \texttt{mgcv} and \texttt{spikeslabgam} have very similar
  predictive performances for the small data sets (\texttt{liver},
  \texttt{pima}), but \texttt{mgcv} is orders of magnitude faster than all of
  the other algorithms.
  \item \texttt{mboost}'s performance is the most stable across replicates on
  each dataset and competitive to the best methods (\texttt{mgcv} and
  \texttt{spikeslabgam}). Its computation time is about the same as that of
  \texttt{spikeslabgam} for the small data sets and three to eight times faster
  for the high-dimensional problems.
  \item \texttt{spikeslabgam} performs consistently
  well on all four data sets, but its computation time is much larger than that
  of its closest competitor on the small data sets (\texttt{mgcv}) and its
  closest competitor on the large data sets (\texttt{mboost}).
  \item \texttt{hgam}'s performance is also fairly stable across replicates for
  the small data sets, but much less competitive.
  For the high-dimensional data sets, it becomes very unstable and does not return
  predictions at all for seven out of ten replicates for \texttt{musk} and no
  predictions at all for \texttt{sonar}.
  \item the performance for \texttt{hypergsplines} is similar but slightly
  worse than \texttt{spikeslabgam} for the smaller problems, but a lack of
  numerical stability of the implementation leads to a failure to return any
  predictions for five of the \texttt{sonar} test sets and six of the
  \texttt{musk} data sets. The remaining fits for
  the high-dimensional problems are not competitive, with relative
  predictive deviance values above 1 for the \texttt{Sonar} dataset indicating
  severe overfitting.
  \item computation times for \texttt{hypergsplines} and \texttt{hgam} are
  much larger than those of the other algorithms which yield better predictions
  in shorter time. The time difference is two orders of magnitude for the small
  datasets.
\end{itemize}

\clearpage

\section{Conclusions}

Currently, function selection is mainly concentrated on additive predictors of
scalar Gaussian or exponential family responses. Extensions to other regression
models for scalar responses, such as survival models, quantile regression and
other beyond mean regression models would be desirable but are still sparse or
completely lacking. A wide and open field for future research is variable and
function selection in multivariate regression and latent variable models, where
the association structure may also depend on covariates.

\section*{Acknowledgements}
Financial support from the German
Science Foundation, grants FA 128/5-1, FA 128/5-2 is gratefully acknowledged. We thank M. Avalos,
H. Liu, and L. Xue for providing software implementing their methods upon
request and D. Saban\'{e}s Bov\'{e} for his generous assistance with the
application of \package{hypergsplines}.

\begin{small}


\begin{thebibliography}{60}
\providecommand{\natexlab}[1]{#1}
\providecommand{\url}[1]{\texttt{#1}}
\expandafter\ifx\csname urlstyle\endcsname\relax
  \providecommand{\doi}[1]{doi: #1}\else
  \providecommand{\doi}{doi: \begingroup \urlstyle{rm}\Url}\fi

\bibitem[Avalos et~al.(2007)Avalos, Grandvalet, and
  Ambroise]{Avalos:Grandvalet:Ambroise:2007}
M.~Avalos, Y.~Grandvalet, and C.~Ambroise.
\newblock {Parsimonious additive models}.
\newblock \emph{Computational Statistics and Data Analysis}, 51:\penalty0
  2851--2870, 2007.

\bibitem[Belitz and Lang(2008)]{Belitz:Lang:2008}
C.~Belitz and S.~Lang.
\newblock {Simultaneous selection of variables and smoothing parameters in
  structured additive regression models}.
\newblock \emph{Computational Statistics and Data Analysis}, 53:\penalty0
  61--81, 2008.

\bibitem[Belitz et~al.(2012)Belitz, Brezger, Kneib, Lang, and Umlauf]{BayesX}
C.~Belitz, A.~Brezger, T.~Kneib, S.~Lang, and N.~Umlauf.
\newblock \emph{\textsf{BayesX} - Software for {B}ayesian inference in
  structured additive regression models}, 2012.
\newblock URL \url{http://www.stat.uni-muenchen.de/~bayesx}.
\newblock Version 2.1.

\bibitem[B{\"u}hlmann and Hothorn(2007)]{Buehlmann:Hothorn:2007}
P.~B{\"u}hlmann and T.~Hothorn.
\newblock {Boosting algorithms: Regularization, prediction and model fitting}.
\newblock \emph{Statistical Science}, 22:\penalty0 477--505, 2007.

\bibitem[B{\"u}hlmann and Yu(2003)]{Buehlmann:Yu:2003}
P.~B{\"u}hlmann and B.~Yu.
\newblock {Boosting with the $l_2$ loss: Regression and Classification}.
\newblock \emph{Journal of the American Statistical Association}, 98:\penalty0
  324--339, 2003.

\bibitem[Cottet et~al.(2008)Cottet, Kohn, and Nott]{Cottet:Kohn:Nott:2008}
R.~Cottet, R.J. Kohn, and D.J. Nott.
\newblock {Variable Selection and Model Averaging in Semiparametric
  Overdispersed Generalized Linear Models}.
\newblock \emph{Journal of the American Statistical Association}, 103:\penalty0
  661--671, 2008.

\bibitem[Eaton et~al.(2008)Eaton, Bateman, and Hauberg]{octave}
J.~W. Eaton, D.~Bateman, and S.~Hauberg.
\newblock \emph{{GNU} \textsf{Octave} Manual Version 3}.
\newblock Network Theory Limited, 2008.

\bibitem[Eilers and Marx(1996)]{EilMar96}
P.~H.~C. Eilers and B.~D. Marx.
\newblock Flexible smoothing using {B}-splines and penalized likelihood.
\newblock \emph{Statistical Science}, 11:\penalty0 89--121, 1996.

\bibitem[Eugster et~al.(2010)Eugster, {Hothorn(Authors)}, Frick, Kondofersky,
  Kuehnle, Lindenlaub, Pfundstein, Speidel, Spindler, Straub, Wickler, and
  {Zink (Contributors)}]{hgam}
M.A. Eugster, T.~{Hothorn(Authors)}, H.~Frick, I.~Kondofersky, O.~S. Kuehnle,
  C.~Lindenlaub, G.~Pfundstein, M.~Speidel, M.~Spindler, A.~Straub, F.~Wickler,
  and K.~{Zink (Contributors)}.
\newblock \emph{\package{hgam}: High-Dimensional Additive Modelling}, 2010.
\newblock \textsf{R} package version 0.1-0.

\bibitem[Fahrmeir and Kneib(2011)]{Fahrmeir:Kneib:2011}
L.~Fahrmeir and T.~Kneib.
\newblock \emph{{Bayesian smoothing and regression for longitudinal, spatial
  and event history data}}.
\newblock Oxford Statistical Science Series 36, Oxford, 2011.

\bibitem[Fahrmeir et~al.(2010)Fahrmeir, Kneib, and
  Konrath]{Fahrmeir:Kneib:Konrath:2010}
L.~Fahrmeir, T.~Kneib, and S.~Konrath.
\newblock {Bayesian regularization in structured additive regression: A
  unifying perspective on shrinkage, smoothing and predictor selection}.
\newblock \emph{Statistics and Computing}, 20:\penalty0 203--219, 2010.

\bibitem[Fan and Li(2001)]{Fan:Li:2001}
J.~Fan and R.~Li.
\newblock {Variable selection via nonconcave penalized likelihood and its
  oracle properties}.
\newblock \emph{Journal of the American Statistical Association}, 96:\penalty0
  1348--1360, 2001.

\bibitem[Frank and Asuncion(2010)]{Frank:Asuncion:2010}
A.~Frank and A.~Asuncion.
\newblock {UCI} machine learning repository, 2010.
\newblock URL \url{http://archive.ics.uci.edu/ml}.

\bibitem[George and McCulloch(1993)]{George:McCulloch:1993}
E.I. George and R.E. McCulloch.
\newblock {Variable selection via Gibbs sampling}.
\newblock \emph{Journal of the American Statistical Association}, 88:\penalty0
  881--889, 1993.

\bibitem[George and McCulloch(1997)]{George:McCulloch:1997}
E.I. George and R.E. McCulloch.
\newblock {Approaches for Bayesian Variable Selection}.
\newblock \emph{Statistica Sinica}, 7:\penalty0 339--374, 1997.

\bibitem[Griffin and Brown(2005)]{Griffin:Brown:05}
J.E. Griffin and P.J. Brown.
\newblock Alternative prior distributions for variable selection with very many
  more variables than observations.
\newblock Technical Report UKC/IMS/05/08, IMS, University of Kent, 2005.

\bibitem[Gu(2002)]{Gu:2002}
C.~Gu.
\newblock \emph{Smoothing Spline ANOVA Models}.
\newblock Springer-Verlag, 2002.

\bibitem[Hothorn et~al.(2012)Hothorn, B{\"u}hlmann, Kneib, Schmid, and
  Hofner]{mboost}
T.~Hothorn, P.~B{\"u}hlmann, T.~Kneib, M.~Schmid, and B.~Hofner.
\newblock \emph{\package{mboost}: Model-Based Boosting}, 2012.
\newblock \textsf{R} package version 2.1-1.

\bibitem[Huang et~al.(2010)Huang, Horowitz, and Wei]{Huang:2010}
J.~Huang, J.L. Horowitz, and F.~Wei.
\newblock {Variable selection in nonparametric additive models}.
\newblock \emph{Annals of Statistics}, 38:\penalty0 2282--2313, 2010.

\bibitem[Ishwaran and Rao(2005)]{Ishwaran:2005}
H.~Ishwaran and J.S. Rao.
\newblock {Spike and slab variable selection: frequentist and Bayesian
  strategies}.
\newblock \emph{Annals of Statistics}, 33\penalty0 (2):\penalty0 730--773,
  2005.

\bibitem[Kneib et~al.(2009)Kneib, Hothorn, and Tutz]{Kneib:Hothorn:Tutz:2009}
T.~Kneib, T.~Hothorn, and G.~Tutz.
\newblock {Variable selection and model choice in geoadditive regression
  models}.
\newblock \emph{Biometrics}, 65:\penalty0 626--634, 2009.

\bibitem[Kneib et~al.(2011)Kneib, Konrath, and Fahrmeir]{KneKonFah09}
T.~Kneib, S.~Konrath, and L.~Fahrmeir.
\newblock High-dimensional structured additive regression models: Bayesian
  regularisation, smoothing and predictive performance.
\newblock \emph{Applied Statistics}, 60:\penalty0 51--70, 2011.

\bibitem[Konrath et~al.(2012)Konrath, Fahrmeir, and
  Kneib]{Konrath:Fahrmeir:Kneib:2012}
S.~Konrath, L.~Fahrmeir, and T.~Kneib.
\newblock Bayesian smoothing, shrinkage and variable selection in hazard
  regression.
\newblock 2012.

\bibitem[Leng and Zhang(2006)]{Leng:Zhang:2006}
C.~Leng and H.H. Zhang.
\newblock {Model selection in nonparametric hazard regression}.
\newblock \emph{Nonparametric Statistics}, 18:\penalty0 417--429, 2006.

\bibitem[Lin and Zhang(2006)]{Lin:Zhang:2006}
Y.~Lin and H.H. Zhang.
\newblock {Component selection and smoothing in multivariate nonparametric
  regression}.
\newblock \emph{Annals of Statistics}, 34:\penalty0 2272--2297, 2006.

\bibitem[Marra and Wood(2011)]{Marra:Wood:2011}
G.~Marra and S.~Wood.
\newblock {Practical variable selection for generalized additive models}.
\newblock \emph{Computational Statistics and Data Analysis}, 55:\penalty0
  2372--2387, 2011.

\bibitem[MATLAB(2010)]{matlab}
MATLAB.
\newblock \emph{\textsf{MATLAB} version 7.10.0 (R2010a)}.
\newblock The MathWorks Inc., Natick, Massachusetts, 2010.

\bibitem[Meier(2009)]{grplasso}
L.~Meier.
\newblock \emph{\package{grplasso}: Fitting user specified models with Group
  Lasso penalty}, 2009.
\newblock \textsf{R} package version 0.4-2.

\bibitem[Meier et~al.(2008)Meier, van~de Geer, and
  B{\"u}hlmann]{Meier:Geer:Buehlmann:2008}
L.~Meier, S.~van~de Geer, and P.~B{\"u}hlmann.
\newblock {The group Lasso for logistic regression}.
\newblock \emph{Journal of the Royal Statistical Society Series B},
  70:\penalty0 53--71, 2008.

\bibitem[Meier et~al.(2009)Meier, van~der Geer, and
  B{\"u}hlmann]{Meier:Geer:Buehlmann:2009}
L.~Meier, S.~van~der Geer, and P.~B{\"u}hlmann.
\newblock {High-dimensional additive modeling}.
\newblock \emph{Annals of Statistics}, 37:\penalty0 3779--3821, 2009.

\bibitem[O'Hara and Sillanp{\"a}{\"a}(2009)]{OHara:Sillanpaa:2009}
R.B. O'Hara and M.J. Sillanp{\"a}{\"a}.
\newblock {A Review of Bayesian Variable Selection Methods: What, How, and
  Which?}
\newblock \emph{Bayesian Analysis}, 4:\penalty0 85--118, 2009.

\bibitem[Panagiotelis and Smith(2008)]{Panagiotelis:Smith:2008}
A.~Panagiotelis and M.~Smith.
\newblock {Bayesian identification, selection and estimation of semiparametric
  functions in high-dimensional additive models}.
\newblock \emph{Journal of Econometrics}, 143:\penalty0 291--316, 2008.

\bibitem[Park and Casella(2008)]{ParCas08}
T.~Park and G.~Casella.
\newblock The {B}ayesian lasso.
\newblock \emph{Journal of the American Statistical Association}, 103:\penalty0
  681--686, 2008.

\bibitem[Polson and Scott(2012)]{Polson:Scott:2012}
N.~G. Polson and J.~G. Scott.
\newblock Local shrinkage rules, {L{\'e}vy} processes and regularized
  regression.
\newblock \emph{Journal of the Royal Statistical Society Series B}, 74\penalty0
  (2):\penalty0 287--311, 2012.

\bibitem[{R Development Core Team}(2011)]{R}
{R Development Core Team}.
\newblock \emph{\textsf{R}: A Language and Environment for Statistical
  Computing}.
\newblock R Foundation for Statistical Computing, Vienna, Austria, 2011.
\newblock URL \url{http://www.R-project.org/}.

\bibitem[Radchenko and James(2010)]{Radchenko:James:2010}
P.~Radchenko and G.M. James.
\newblock {Variable selection using adaptive nonlinear interaction structures
  in high dimensions}.
\newblock \emph{Journal of the American Statistical Association}, 105:\penalty0
  1--13, 2010.

\bibitem[Ravikumar et~al.(2009)Ravikumar, Liu, Lafferty, and
  Wasserman]{Ravikumar:2009}
P.~Ravikumar, H.~Liu, J.~Lafferty, and L.~Wasserman.
\newblock {Sparse additive models}.
\newblock \emph{Journal of the Royal Statistical Society Series B},
  71:\penalty0 1009--1030, 2009.

\bibitem[Reich et~al.(2009)Reich, Storlie, and
  Bondell]{Reich:Storlie:Bondell:2009}
B.J. Reich, C.B. Storlie, and H.D. Bondell.
\newblock {Variable selection in {B}ayesian smoothing spline ANOVA models:
  Application to deterministic computer codes}.
\newblock \emph{Technometrics}, 51:\penalty0 110, 2009.

\bibitem[Rue and Held(2005)]{RueHel05}
H.~Rue and L.~Held.
\newblock \emph{Gaussian {M}arkov Random Fields}.
\newblock Chapman \& Hall / CRC, 2005.

\bibitem[{Saban{\'e}s Bov{\'e}}(2012)]{hypergsplines}
D.~{Saban{\'e}s Bov{\'e}}.
\newblock \emph{\package{hypergsplines}: {B}ayesian model selection with
  penalised splines and hyper-g prior}, 2012.
\newblock \textsf{R} package version 0.0-32.

\bibitem[{Saban{\'e}s Bov{\'e}} et~al.(2011){Saban{\'e}s Bov{\'e}}, Held, and
  Kauermann]{Sabanes:Held:Kauermann:2011}
D.~{Saban{\'e}s Bov{\'e}}, L.~Held, and G.~Kauermann.
\newblock {Mixtures of g-Priors for Generalised Additive Model Selection with
  Penalised Splines}.
\newblock Technical report, University of Zurich and University Bielefeld,
  2011.
\newblock URL \url{http://arxiv.org/abs/1108.3520}.

\bibitem[Scheipl(2011{\natexlab{a}})]{Scheipl:2011}
F.~Scheipl.
\newblock \emph{Bayesian Regularization and Model Choice in Structured Additive
  Regression}.
\newblock PhD thesis, Ludwig-Maximilians-Universit{\"a}t M{\"u}nchen,
  2011{\natexlab{a}}.

\bibitem[Scheipl(2011{\natexlab{b}})]{spikeSlabGAM}
F.~Scheipl.
\newblock \texttt{spikeSlabGAM}: {B}ayesian variable selection, model choice
  and regularization for generalized additive mixed models in \textsf{R}.
\newblock \emph{Journal of Statistical Software}, 43\penalty0 (14):\penalty0
  1--24, 9 2011{\natexlab{b}}.
\newblock URL \url{http://www.jstatsoft.org/v43/i14}.

\bibitem[Scheipl et~al.(2012)Scheipl, Fahrmeir, and
  Kneib]{Scheipl:Fahrmeir:Kneib:2011}
F.~Scheipl, L.~Fahrmeir, and T.~Kneib.
\newblock Spike-and-slab priors for function selection in structured additive
  regression models.
\newblock \emph{Journal of the American Statistical Association}, 2012.
\newblock URL \url{http://arxiv.org/abs/1105.5250}.
\newblock in press.

\bibitem[Smith and Kohn(1996)]{Smith:Kohn:1996}
M.~Smith and R.~Kohn.
\newblock {Nonparametric regression using Bayesian variable selection}.
\newblock \emph{Journal of Econometrics}, 75:\penalty0 317--344, 1996.

\bibitem[Storlie et~al.(2011)Storlie, Bondell, Reich, and Zhang]{Storlie:2011}
C.~Storlie, H.~Bondell, B.~Reich, and H.H. Zhang.
\newblock {Surface estimation, variable selection, and the nonparametric oracle
  property}.
\newblock \emph{Statistica Sinica}, 21\penalty0 (2):\penalty0 679--705, 2011.

\bibitem[Tibshirani(1996)]{Tibshirani:1996}
R.~Tibshirani.
\newblock {Regression shrinkage and selection via the Lasso}.
\newblock \emph{Journal of the Royal Statistical Society Series B},
  58:\penalty0 267--288, 1996.

\bibitem[Tutz and Binder(2006)]{Tutz:Binder:2006}
G.~Tutz and H.~Binder.
\newblock {Generalized additive modelling with implicit variable selection by
  likelihood based boosting}.
\newblock \emph{Biometrics}, 62:\penalty0 961--971, 2006.

\bibitem[Umlauf et~al.(2012)Umlauf, Kneib, and Lang]{R2BayesX}
N.~Umlauf, T.~Kneib, and S.~Lang.
\newblock \emph{\package{R2BayesX}: Estimate Structured Additive Regression
  Models with {BayesX}}, 2012.
\newblock \textsf{R} package Version 0.1-1.

\bibitem[Wahba(1990)]{Wahba:1990}
G.~Wahba.
\newblock \emph{Spline Models for Observational Data}.
\newblock SIAM, 1990.

\bibitem[Wang et~al.(2007)Wang, Chen, and Li]{Wang:Chen:Li:2007}
L.~Wang, G.~Chen, and H.~Li.
\newblock {Group SCAD regression analysis for microarray time course gene
  expression data}.
\newblock \emph{Bioinformatics}, 23:\penalty0 1486--1494, 2007.

\bibitem[Wood(2012)]{mgcv}
S.~Wood.
\newblock \emph{\package{mgcv}: GAMs with GCV/AIC/REML smoothness estimation
  and GAMMs by PQL}, 2012.
\newblock \textsf{R} package version 1.7-18.

\bibitem[Wood et~al.(2002)Wood, Kohn, Shively, and Jiang]{Wood:Kohn:2002}
S.~Wood, R.~Kohn, T.~Shively, and W.~Jiang.
\newblock {Model selection in spline nonparametric regression}.
\newblock \emph{Journal of the Royal Statistical Society Series B},
  64:\penalty0 119--139, 2002.

\bibitem[Xue(2009)]{Xue:2009}
L.~Xue.
\newblock {Consistent variable selection in additive models}.
\newblock \emph{Statistica Sinica}, 19:\penalty0 1281--1296, 2009.

\bibitem[Yau et~al.(2003)Yau, Kohn, and Wood]{Yau:Kohn:Wood:2003}
P.~Yau, R.~Kohn, and S.~Wood.
\newblock {Bayesian variable selection and model averaging in high-dimensional
  multinomial nonparametric regression}.
\newblock \emph{Journal of Computational and Graphical Statistics},
  12:\penalty0 23--54, 2003.

\bibitem[Yuan and Lin(2006)]{Yuan:Lin:2006}
M.~Yuan and Y.~Lin.
\newblock {Model selection and estimation in regression with grouped
  variables}.
\newblock \emph{Journal of the Royal Statistical Society Series B},
  68:\penalty0 49--67, 2006.

\bibitem[Zhang et~al.(2011{\natexlab{a}})Zhang, Cheng, and Liu]{zhang2011}
H.~Zhang, G.~Cheng, and Y.~Liu.
\newblock Linear or nonlinear? automatic structure discovery for partially
  linear models.
\newblock \emph{Journal of the American Statistical Association}, 106:\penalty0
  1099--1112, 2011{\natexlab{a}}.

\bibitem[Zhang and Lin(2006)]{Zhang:Lin:2006}
H.H. Zhang and Y.~Lin.
\newblock {Component selection and smoothing for nonparametric regression in
  exponential families}.
\newblock \emph{Statist. Sinica}, 16:\penalty0 1021--1041, 2006.

\bibitem[Zhang et~al.(2011{\natexlab{b}})Zhang, Cheng, and
  Liu]{Zhang:Cheng:Liu:2011}
H.H. Zhang, G.~Cheng, and Y.~Liu.
\newblock Linear or nonlinear? automatic structure discovery for partially
  linear models.
\newblock \emph{Journal of the American Statistical Association}, 106\penalty0
  (495):\penalty0 1099--1112, 2011{\natexlab{b}}.

\bibitem[Zou(2006)]{Zou:2006}
H.~Zou.
\newblock {The adaptive Lasso and its oracle properties}.
\newblock \emph{Journal of the American Statistical Association}, 101:\penalty0
  1418--1429, 2006.

\end{thebibliography}

\clearpage

\end{small}

\end{document}